\documentclass[journal]{IEEEtran}
\usepackage{amsmath}
\usepackage{amsfonts}
\usepackage{amssymb}
\usepackage{algorithmic}
\usepackage{array}
\usepackage{url}
\usepackage{microtype}
\usepackage{graphicx}
\usepackage{subfigure}
\usepackage{booktabs} % for professional tables
\usepackage{tabularx, enumitem}
\usepackage{hyperref}
\usepackage{cite}
\newcommand{\argminA}{\arg\,\min}
\newcommand\numberthis{\addtocounter{equation}{1}\tag{\theequation}}
\usepackage[ruled, vlined]{algorithm2e}
\DontPrintSemicolon
\usepackage{multirow}
\usepackage{setspace}

\usepackage{color,soul}
\newcommand{\rgbsymbol}{\includegraphics[height=1.5ex]{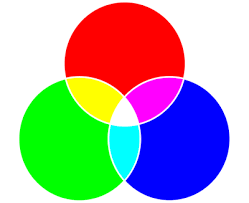}}
\begin{document}

\title{{\it Nu}SPAN: A Proximal Average Network for Nonuniform Sparse Model --- Application to Seismic Reflectivity Inversion}

\author{Swapnil~Mache$^\dagger$,
        Praveen~Kumar~Pokala$^\dagger$,
        Kusala~Rajendran,
        and~Chandra~Sekhar~Seelamantula,~\IEEEmembership{Senior Member,~IEEE}% <-this % stops a space
\thanks{$^\dagger$ Equal contribution.}%
\thanks{Swapnil Mache is with the Centre of Excellence in Advanced Mechanics of Materials, Indian Institute of Science (IISc), Bangalore, India, and with the Department of Electrical Engineering, IISc. He was formerly with the Centre for Earth Sciences, IISc. Email:~machesanjay@iisc.ac.in.}%
\thanks{Praveen Kumar Pokala and Chandra Sekhar Seelamantula are with the Department of Electrical Engineering, IISc. E-mail:~praveenkumar.pokala@gmail.com, chandra.sekhar@ieee.org. }%
\thanks{Kusala Rajendran is with the Centre of Excellence in Advanced Mechanics of Materials, IISc. She is Retired Professor from the Centre for Earth Sciences, IISc. Email:~kusalaraj@gmail.com.}
}

\maketitle

\begin{abstract}
    We solve the problem of sparse signal deconvolution in the context of seismic reflectivity inversion, which pertains to high-resolution recovery of the subsurface reflection coefficients. Our formulation employs a nonuniform, non-convex synthesis sparse model comprising a combination of convex and non-convex regularizers, which results in accurate approximations of the $\ell_0$ pseudo-norm. The resulting iterative algorithm requires the proximal average strategy. When unfolded, the iterations give rise to a learnable proximal average network architecture that can be optimized in a data-driven fashion. We demonstrate the efficacy of the proposed approach through numerical experiments on synthetic $1$-D seismic traces and $2$-D wedge models in comparison with the benchmark techniques. We also present validations considering the simulated Marmousi$2$ model as well as real $3$-D seismic volume data acquired from the Penobscot $3$D survey off the coast of Nova Scotia, Canada.
\end{abstract}


\IEEEpeerreviewmaketitle

\section{Introduction}
\begin{figure*}[t]
    \centering
    \resizebox{0.9\linewidth}{!}{
    \includegraphics[width=\columnwidth]{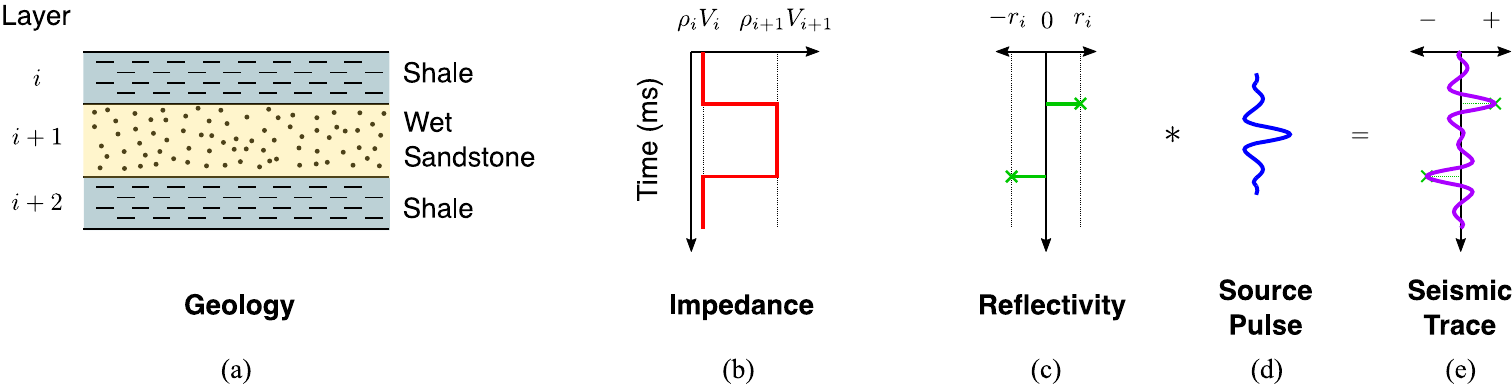}}
    \caption{\protect\rgbsymbol~An ideal three-layer subsurface model. The operator $*$ denotes convolution.}
    \label{fig:model2}
\end{figure*}

\IEEEPARstart{R}{eflectivity} inversion, which is a seismic deconvolution problem in reflection seismology, is the means by which one can characterize and image the layered subsurface structure. A recorded seismic trace $\boldsymbol{y} \in \mathbb{R}^{n}$ is modeled as the convolution of the source pulse $\boldsymbol{h}$, assumed to be a Ricker wavelet \cite{shearer2009introduction}, and the subsurface reflectivity $\boldsymbol{x} \in \mathbb{R}^{n}$, given by \( \boldsymbol{y} = \boldsymbol{h}*\boldsymbol{x} + \boldsymbol{n}, \)
% \begin{equation}\label{convmodel}
%     \boldsymbol{y} = \boldsymbol{h}*\boldsymbol{x} + \boldsymbol{n},
% \end{equation}%
where $\boldsymbol{n}$ is the noise and $*$ denotes convolution. Reflectivity is modeled as a sparse vector comprising subsurface layers, each with a constant impedance, giving rise to a piecewise-constant impedance model \cite{oldenburg1983recovery, yilmaz2001seismic}. Figure~\ref{fig:model2} illustrates the model with two shale layers sandwiching a wet sand layer (Figure~\ref{fig:model2}(a)) \cite{russell2019machine}. The reflectivity (Figure~\ref{fig:model2}(c)) convolved with the source (Figure~\ref{fig:model2}(d)) is recorded at the receiver as a seismic trace (Figure~\ref{fig:model2}(e)).

The subsurface geology (Figure~\ref{fig:model2}(a)) is related to reflectivity (Figure~\ref{fig:model2}(c)) through acoustic impedance (Figure~\ref{fig:model2}(b)), which is defined as the product of density ($\rho$) and P-wave velocity ($V$) \cite{oldenburg1983recovery}, given by
\begin{equation}\label{rhov}
    r_{i} = \frac{\rho_{i+1}V_{i+1} - \rho_{i}V_{i}}{\rho_{i+1}V_{i+1} + \rho_{i}V_{i}}.
\end{equation}%

Solving the ill-posed linear inverse problem of estimating the subsurface reflectivity through the classical least-squares formulation \cite{taylor1979deconvolution} leads to nonuniqueness issues arising out of a convolution with a bandlimited wavelet and loss of low and high-frequency information \cite{oldenburg1983recovery, berkhout1977least, debeye1990lp}. The nonuniqueness aspect can be tackled through regularization \cite{tarantola2005inverse}, for example, by imposing a sparsity prior on the solution through the $\ell_1$-norm \cite{oldenburg1983recovery, taylor1979deconvolution}.

Zhang and Castagna \cite{zhang2011seismic} solved the $\ell_1$-norm constrained reflectivity inversion problem through basis-pursuit inversion (BPI) \cite{chen2001atomic}, using a wavelet dictionary of odd and even reflectivity pairs. The fast iterative shrinkage-thresholding algorithm (FISTA) \cite{beck2009fast} has been employed for reflectivity inversion \cite{perez2012inversion} along with an amplitude recovery boost through debiasing steps of least-squares inversion \cite{perez2013high} and ``adding back the residual'' \cite{li2020debiasing}. The $\ell_1$-norm, although convex, is not the best sparsity constraint for reflectivity inversion, and the accurate estimation of the sparsity of seismic reflections is challenging \cite{li2019optimal, yuan2019seismic}. Further, $\ell_1$-norm regularization underestimates the high-amplitude components and introduces a bias in the estimate of the sparse code $\boldsymbol{x}$ \cite{candes2008enhancing, zhang2010analysis, selesnick2017sparse}. Both \cite{perez2013high} and \cite{li2020debiasing} observed the attenuation of reflectivity magnitudes due to the $\ell_1$-norm regularization term and adopted a post-processing debiasing step \cite{wright2009sparse}.

Non-convex regularization strategies have been adopted to overcome the shortcomings of $\ell_1$ regularization in sparse recovery problems \cite{selesnick2017sparse, zhang2010nearly, woodworth2016compressed}. Non-convex penalties such as the smoothly clipped absolute deviation (SCAD) \cite{fan2001variable} and minimax concave penalty (MCP) \cite{zhang2010nearly} have been shown to be superior to $\ell_1$ and non-convex regularization approaches in the context of inverse problems. In the context of reflectivity inversion, \cite{li2019optimal} introduced a data-driven $\ell_p$-loss-$\ell_q$-regularization ($p=2, 0<q<1$), with an adaptive approach for choosing the optimal $q$. Recently, \cite{zhong2014gradient} and \cite{kamilov2016parallel} developed proximal gradient-descent methods based on the proximal average strategy \cite{bauschke2008proximal, yu2013better}, by considering composite regularization, which is a convex combination of several sparsity-inducing regularizers. 

A new class of deep neural network (DNN) architectures was introduced by \cite{gregor2010learning}, which are inspired by unrolling iterative algorithms into learnable networks \cite{monga2021algorithm}. Gregor and LeCun \cite{gregor2010learning} proposed the learned iterative shrinkage and thresholding algorithm (LISTA), based on unfolding the update steps of ISTA \cite{daubechies2004iterative} into the layers of a neural network. This class of model-based architectures has been demonstrated to be effective in solving sparse linear inverse problems \cite{mahapatra2017deep, mukherjee2017dnns, zhang2017ista, borgerding2017amp, sreter2018learned, liu2019deep, pokala2019firmnet, li2020efficient, shlezinger2020modelbased, pokala2020confirmnet, wang2020dnu, tolooshams2020deep, jawali2020cornet, tolooshams2021unfolding}. Recently, Monga {\it et al.} \cite{monga2021algorithm} reviewed algorithm unrolling and its applications to imaging, vision and recognition, speech processing, and other signal and image processing problems. 

Recently, learning-based approaches have also been employed for solving inversion problems in seismology and geophysics \cite{yuan2013spectral, yuan2019seismic, russell2019machine, bergen2019machine, adler2021deep, lewis2017deep, richardson2018seismic, ovcharenko2019deep, sun2020extrapolated, araya2018deep, kim2018geophysical, wang2018velocity, wu2018inversionnet, yang2019deep, adler2019deep, das2019convolutional, park2020automatic, wu2020seismic, wang2020velocity, biswas2019prestack, alfarraj2019semi, araya2019deep, wang2019seismic, mosser2020stochastic, zhang2020data}, including seismic reflectivity inversion \cite{yuan2013spectral, yuan2019seismic, russell2019machine, adler2021deep, kim2018geophysical, biswas2019prestack}. Adler {\it et al.} reviewed the development of deep learning solutions in the context of seismology and geophysics from approaches that enhanced the performance of full waveform inversion \cite{lewis2017deep, richardson2018seismic, ovcharenko2019deep, sun2020extrapolated} to end-to-end seismic inversion \cite{araya2018deep, kim2018geophysical, wang2018velocity, wu2018inversionnet, yang2019deep, adler2019deep, das2019convolutional, park2020automatic, wu2020seismic, wang2020velocity} that employed physics-guided architectures \cite{biswas2019prestack, alfarraj2019semi} and deep generative modeling \cite{araya2019deep, wang2019seismic, mosser2020stochastic, zhang2020data}.

Kim and Nakata \cite{kim2018geophysical} employed an elementary feedforward neural network and observed superior support recovery compared to a least-squares approach but poor amplitude recovery. Yuan and Su \cite{yuan2019seismic}, and earlier, \cite{yuan2013spectral}, employed sparse Bayesian learning (SBL) for recovering the sparse reflection coefficients by maximizing the marginal likelihood either by using the expectation-maximization algorithm (SBL-EM) \cite{wipf2004sparse, yuan2019seismic}, or by sequentially updating the sparsity-controlling hyperparameters through a sequential algorithm-based approach \cite{yuan2013spectral}. The former \cite{wipf2004sparse, yuan2019seismic} was demonstrated as being more robust to noise, having higher accuracy, and preserving lateral continuity in the seismic profile. 

\subsection{Motivation and Contribution}
Estimating the sparsity of the reflection data is difficult \cite{li2019optimal, yuan2019seismic} and the recovery of support is prioritized \cite{shearer2009introduction}. One can resort to data-driven approaches, which outperform conventional techniques, especially in support recovery \cite{kim2018geophysical}, when the knowledge about the underlying geology is limited, with an added computational benefit \cite{russell2019machine}. Further, as opposed to an elementary feedforward neural network approach for seismic reflectivity inversion \cite{kim2018geophysical}, one can employ model-based learning frameworks \cite{shlezinger2020modelbased} such as deep-unrolled architectures \cite{monga2021algorithm}. Such deep neural networks, where the architecture is informed by the inverse problem itself, can provide insights into the problem as well as model interpretability that is critical to gaining physical insights into the system under consideration \cite{bergen2019machine}. In our previous work \cite{mache2021durin}, we demonstrated the efficacy of FirmNet \cite{pokala2019firmnet} and LISTA-like \cite{gregor2010learning} formulations in solving the seismic reflectivity inversion problem, comparing the approaches with baselines such as BPI \cite{chen2001atomic, zhang2011seismic}, FISTA \cite{beck2009fast, perez2012inversion}, and SBL-EM \cite{wipf2004sparse, yuan2019seismic}. Here, we expand the work by constructing and learning from the data \cite{bergen2019machine}, a composite nonuniform sparse prior from a convex combination of weighted counterparts of three sparsity-promoting penalties (the $\ell_1$ norm, MCP \cite{zhang2010nearly}, and SCAD \cite{fan2001variable} penalties).

The contributions of this paper are stated below.
\begin{enumerate}
    \item We consider the problem of seismic reflectivity inversion based on a data-driven prior, as opposed to a pre-designated prior, within the framework of deep-unfolding. To the best of our knowledge, data-driven priors have not been explored for solving the problem of seismic reflectivity inversion.
    \item We propose an optimization framework for seismic reflectivity inversion based on a composite sparse-prior, which comprises multiple weighted regularizers. The weights are allowed to be different for each component. Such a model is commonly referred to as a {\it nonuniform sparse model}.
    \item We develop the nonuniform proximal-averaged thresholding algorithm ({\it Nu}PATA) and its deep-unfolded version, the nonuniform sparse proximal average network ({\it Nu}SPAN) to solve the problem under consideration.
    \item We demonstrate the efficacy of the proposed network, {\it Nu}SPAN, over synthetic $1$-D and $2$-D datasets, and simulated and real datasets.
\end{enumerate}

\subsection{Organization of the Paper and Notations}
This paper is structured as follows. In Section~\ref{sec:nu_span1}, we introduce the nonuniform sparse model for sparse seismic reflectivity inversion. We develop an optimization algorithm and the corresponding unrolled network for solving the problem under consideration. Section~\ref{sec:nu_span2} explains a generalized variant of the problem formulation discussed in Section~\ref{sec:nu_span1}. In Section~\ref{sec:results}, we present experimental results and demonstrate the efficacy of the proposed proximal average network ({\it Nu}SPAN) in comparison with baselines. Conclusions are provided in Section~\ref{sec:conclusions}.

The notational conventions used in this paper are listed in Table~\ref{table:notation}.

\begin{table}[t]
    \centering
    \caption{Notations}
    \label{table:notation}
    \resizebox{\linewidth}{!}{
    \begin{tabular}{c|l}
        \toprule
        \bfseries Notation & \bfseries Description \\
        \midrule
        $x$ & Scalar (lower-case letter) \\
        $\boldsymbol{x}$ & Vector (lower-case bold letter) \\
        ${\left\| \boldsymbol{x} \right\|}_{p}$ & $\ell_p-\mathrm{norm}$ of $\boldsymbol{x}$ \\
        $\boldsymbol{x}^{(k)}$ & $\boldsymbol{x}$ at the $k^{th}$ iteration of an algorithm \\
        $x_i$ & $i^{th}$ element of $\boldsymbol{x}$ \\
        $\boldsymbol{x}'$ & Flipped version of $\boldsymbol{x}$ \\
        $\boldsymbol{1}$ & Vector of all ones \\
        $\mathbf{H}$ & Matrix (upper-case bold letter) \\
        $\mathbf{H}^{\dagger}$ & Pseudo-inverse of $\mathbf{H}$ \\
        $\mathbf{I}$ & Identity matrix \\
        $\mathrm{sgn}$ & Signum function \\
        $\nabla$ & Gradient operator \\
        $*$ & Convolution operator \\
        $\langle \cdot, \cdot \rangle$ & Inner-product \\
        $\oplus$ & Element-wise sum \\
        $\odot$ & Element-wise product \\
        $\mathbb{R}^{n}_{> t}$ & Set of vectors in $\mathbb{R}^{n}$ with entries greater than $t$ \\
        \bottomrule
    \end{tabular}}
\end{table}

\begin{table*}[t]
    \centering
    \caption{Sparsity-promoting regularizers used in this study and corresponding proximal operators.}
    \label{table:regularizers}
    \resizebox{\linewidth}{!}{
    \begin{tabular}{p{0.21\linewidth}|c|c}
        \toprule
        Name & Penalty Function & Proximal Operator\\
        \midrule
        $\ell_1-\mathrm{norm}$ $(\lambda>0)$ & $g_1(x)=\lambda \left| x \right|$ & $\mathcal{P}_{g_1}(x) = \mathrm{sgn}(x) ~ \mathrm{max}\left(\left| x \right| - \lambda, 0\right)$\\
        \midrule
        MCP \cite{zhang2010nearly} $(\mu>0, \gamma>1)$ &%
        $g_2(x)=\begin{cases} {\mu \left( \left| x \right| - \cfrac{{\left| x \right|}^2}{2 \mu \gamma } \right) , } \quad {\mathrm{for} \left| x \right| \leq \gamma\mu ,} \\
        {\cfrac{{\mu }^{2}\gamma }{2}, } \qquad\qquad\qquad\, {\mathrm{for} \left| x \right| \geq \gamma \mu .} \end{cases}$ &%
        $\mathcal{P}_{g_2}(x)=\begin{cases} {0,} \qquad\qquad\qquad\qquad\qquad {\mathrm{for} \; \left| x \right| \leq \mu,} \\%[5pt]
        {\mathrm{sgn} \left(x\right) \cfrac{\gamma}{\gamma - 1} \left( \left| x \right| - \mu \right),} \enspace\, {\mathrm{for} \; \mu < \left| x \right| \leq \gamma \mu,} \\%[5pt]
        {x,} \qquad\qquad\qquad\qquad\qquad {\mathrm{for} \; \left| x \right| > \gamma \mu .} \end{cases}$ \\%
        \midrule
        SCAD \cite{fan2001variable} $(\nu>0, a>2)$ &%
        $g_3(x)=\begin{cases} {\nu \left| x \right|,} \qquad\qquad\qquad\quad\, {\mathrm{for} \left| x \right| \leq \nu ,} \\%[5pt]
        {\cfrac{{\left| x \right|}^{2} - 2a\nu \left| x \right| + \nu ^{2}}{2(1-a)},} \enspace\,\, {\mathrm{for}  \nu < \left| x \right| \leq a\nu ,} \\%[5pt]
        {\cfrac{(a+1){\nu ^{2}}}{2},} \qquad\qquad\enspace\,\, {\mathrm{for} \left| x \right| > a\nu .} \end{cases}$ &%
        $\mathcal{P}_{g_3}(x)=\begin{cases} {\mathrm{sgn}\left(x\right) \mathrm{max} \left( \left| x \right| - \nu, 0 \right),} \, {\mathrm{for} \; \left| x \right| \leq 2\nu,} \\%[5pt]
        {\cfrac{\left(a - 1\right)x - \mathrm{sgn} \left(x\right) a \nu}{a-2},} \enspace\, {\mathrm{for} \; 2\nu < \left| x \right| \leq a \nu,} \\%[5pt]
        {x,} \qquad\qquad\qquad\qquad\qquad {\mathrm{for} \; \left| x \right| > a \nu .} \end{cases}$ \\
        \bottomrule
    \end{tabular}}
\end{table*}

\section{Non-convex penalties and their weighted counterparts}
\label{sec:penalties}
\begin{figure}[t]
    \centering
    \includegraphics[width=\linewidth]{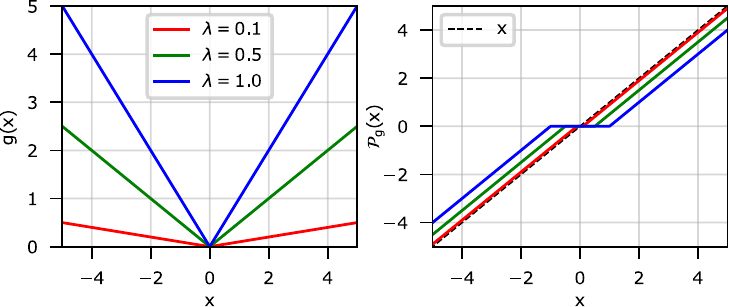}
    \caption{\protect\rgbsymbol~The $\ell_1$ penalty and its corresponding proximal operator, the soft-thresholding operator.} \label{fig:l1}
\end{figure}

\begin{figure}[t]
    \centering
    \includegraphics[width=\linewidth]{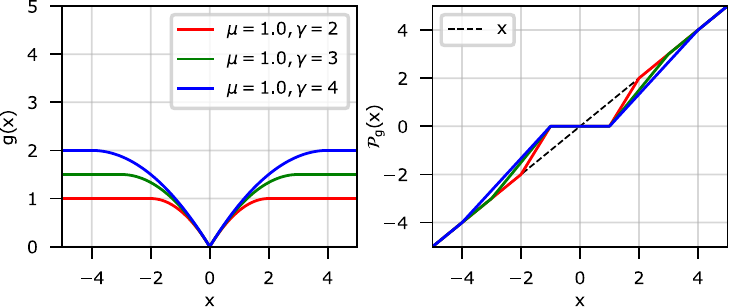}
    \caption{\protect\rgbsymbol~The minimax concave penalty (MCP) and its corresponding proximal operator, the firm-thresholding operator.} \label{fig:mcp}
\end{figure}

\begin{figure}[t]
    \centering
    \includegraphics[width=\linewidth]{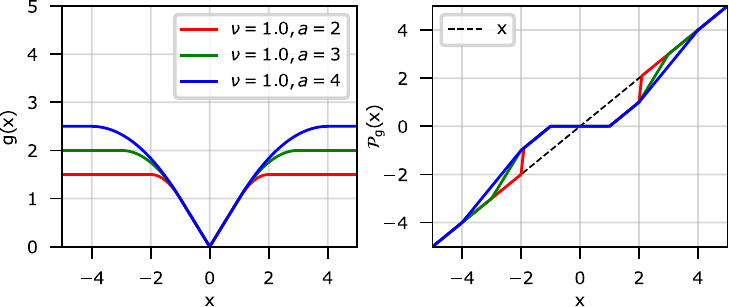}
    \caption{\protect\rgbsymbol~The smoothly clipped absolute deviation (SCAD) penalty and its corresponding proximal operator.} \label{fig:scad}
\end{figure}

The $\ell_1$ penalty (Fig.~\ref{fig:l1}), which is the convex relaxation of the $\ell_0$ penalty, is the most popular one for promoting sparsity while solving sparse linear inverse problems, owing to its convexity-preserving attribute, making the optimization problem computationally more tractable than $\ell_0$ or non-convex regularization. However, $\ell_1$-regularization results in a biased estimate. In particular, it over-penalizes large-valued coefficients due to the constant shrinkage parameter. Non-convex penalties such as the minimax concave penalty (MCP) (Fig.~\ref{fig:mcp}) and smoothly clipped absolute deviation (SCAD) (Fig.~\ref{fig:scad}) result in nearly unbiased estimates compared to the $\ell_1$ penalty. Non-convex regularization overcomes the limitation of $\ell_1$, but the resulting optimization problem is quite challenging. Proximal operators are central to developing proximal gradient-descent methods for optimizing non-convex, nonsmooth sparse linear inverse problems. Further, the resulting optimization problem is solved via proximal methods since proximal operators of MCP and SCAD are computationally simple \cite{zhang2010analysis, fan2001variable}. The proximal operator $\mathcal{P}_g:\mathbb{R} \to \mathbb{R}$ associated with the penalty $g: \mathbb{R} \to \mathbb{R}_{\geq 0}$ is defined as:
\begin{align}
    \mathcal{P}_g(y,\theta_g)=\argminA_{x}\frac{1}{2}(x-y)^{2}+\lambda g(x),
\end{align}
where $\theta_g$ denotes the parameters of $g$ and $\lambda > 0$.

The scaled $\ell_1$ penalty is defined as 
\begin{equation}
    \label{eq:l1}
    g_1(x; \lambda)=\lambda \left| x \right|,
\end{equation}
where the weight factor $\lambda$ allows for adjusting the threshold value. The MCP $g_{2}: \mathbb{R} \rightarrow \mathbb{R}_{\geq 0}$ is defined as
\begin{gather*}
	g_{2}(x; \mu, \gamma) = \left\{\begin{array}{lr}
		\displaystyle \mu |x| - \frac{|x|^2}{2\gamma}, & |x| \leq \mu\gamma,\\
		\displaystyle \frac{\mu^2 \gamma}{2}, & |x| \geq \mu\gamma,
	\end{array}\right. \numberthis \label{MCP}
\end{gather*}
where $\mu > 0, \gamma > 1$. The SCAD penalty $g_{3}: \mathbb{R} \rightarrow \mathbb{R}_{\geq 0}$ is defined as
\begin{gather*}
	g_{3}(x;\nu,a) = \left\{\begin{array}{lr}
		\nu|x|, & |x|\leq\nu,\\
		\displaystyle \frac{2\nu a|x|-|x|^{2}-\nu^{2}}{2(a-1)}, & \nu<|x|\leq \nu a,\\
		\displaystyle \frac{\nu^{2}(a+1)}{2}, & |x|>\nu a,
	\end{array}\right. 
	\numberthis \label{SCAD}
\end{gather*}
where $\nu > 0, a > 2$. The penalties used in this study and their proximal operators are shown in Table~\ref{table:regularizers}. 

\begin{figure*}[!t]
    \vskip 0.2in
    \centering
    \resizebox{\linewidth}{!}{
    \includegraphics[width=\columnwidth]{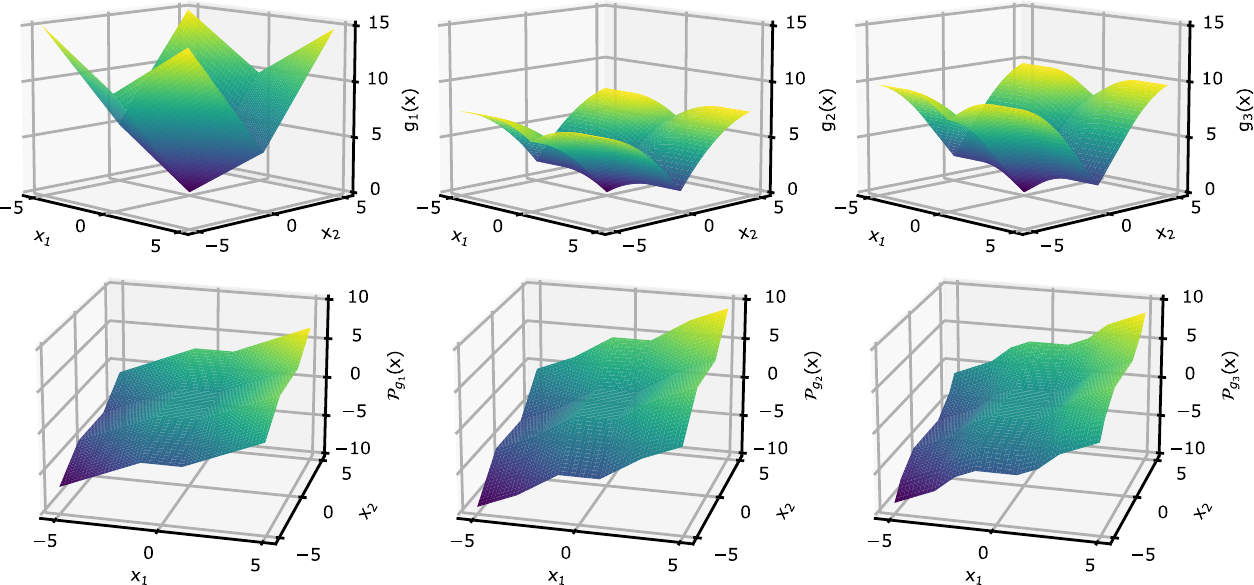}}
    \caption{\protect\rgbsymbol~The penalty functions and proximal operators corresponding to Table~\ref{table:regularizers}.}
    \label{fig:regularizers}
\end{figure*}

We propose weighted counterparts of the above regularizers as atoms to construct a learnable sparsity-prior. Further, the chosen penalty functions have closed-form proximal operators. The weighted counterparts of the penalty functions and their proximal operators are given in Figure~\ref{fig:regularizers}.

\begin{enumerate}
    \item $g_{1}(\boldsymbol{x}) = \Vert\boldsymbol{\lambda} \odot \boldsymbol{x}\Vert_1$, where $\boldsymbol{\lambda} \in \mathbb{R}^{n}_{> 0}$, and $\odot$ denotes element-wise product. $\mathbb{R}^{n}_{> 0}$ denotes the set of vectors in $\mathbb{R}^{n}$ containing positive entries.
    \item $g_2(\boldsymbol{x})$ is weighted MCP defined as $g_2(\boldsymbol{x}) = \sum_{j=1}^n g_2(x_j).$
    \item $g_3(\boldsymbol{x})$ is weighted SCAD, which is defined as $g_3(\boldsymbol{x}) = \sum_{j=1}^n g_3(x_j).$
\end{enumerate}

% The parameters of $g$ are learned in a data-driven setting, with the trainable parameters of the proposed sparsity-prior being $\{\boldsymbol{\lambda, \mu, \nu}\} \in \mathbb{R}^n_{>0}$ and $\{\boldsymbol{\gamma}\in \mathbb{R}^n_{>1}, \boldsymbol{a} \in \mathbb{R}^n_{>2} \}$. $\mathbb{R}^{n}_{> 1}$ denotes the set of vectors in $\mathbb{R}^{n}$ with entries greater than $1$.

\section{Proximal Average Network for Reflectivity Inversion --- Nonuniform Sparse Model (Type-\texorpdfstring{$1$}{1})}
\label{sec:nu_span1}

The sparse reflection coefficients $\boldsymbol{x} \in \mathbb{R}^n$ are recovered from the observed noisy seismic trace $\boldsymbol{y} \in \mathbb{R}^n$ via optimization of the following composite-regularized cost:
\begin{equation}\label{nu_comp_reg1}
\begin{split}
    \arg\min_{\boldsymbol{x}, \, \{\omega_{i}\}} \, \Big\{\mathcal{J(\boldsymbol{x})} = \underbrace{\frac{1}{2}{\left\| \boldsymbol{h}*\boldsymbol{x}-\boldsymbol{y} \right\|}_{2}^{2}}_{f(\boldsymbol{x})} + \underbrace{\sum\limits_{i = 1}^m \omega_{i}{ g_{i}(\boldsymbol{x}) }}_{g(\boldsymbol{x})}\Big\} \\
    \text{subject to} \, \, \sum\limits_{i=1}^m {\omega_{i}=1}, 0<\omega_{i}<1 \; \forall{i},
\end{split}
\end{equation}%
where $m$ is the number of regularizers, and $\{\omega_{i}\}$ are the weights assigned to the regularizers.

The composite objective function $\mathcal{J}$ in \eqref{nu_comp_reg1}, regularizer $g$ and data-fidelity term $f$ satisfy the following properties, which are essential for minimizing $\mathcal{J}$,
\begin{enumerate}
	\item[P1.] $f : \mathbb{R}^{n} \rightarrow \left (- \infty,\infty \right] $ is proper, closed, and $L$-smooth, i.e., $\Vert \nabla f(\boldsymbol{x}) - \nabla f(\boldsymbol{y})\Vert_{2} \leq L \Vert \boldsymbol{x} - \boldsymbol{y}\Vert_{2}$.
	\item[P2.] $g$ is lower semi-continuous.
	\item[P3.] $\mathcal{J}$ is bounded from below, i.e., inf $\mathcal{J} > -\infty$.
\end{enumerate}

Given a set of training samples $\{\boldsymbol{y}_p, \boldsymbol{x}_p\}, \, \, \forall p \in \{1, 2, 3, \ldots , N\}$, our objective is to design a deep-unrolled network that solves the optimization problem stated in Eq.~\eqref{nu_comp_reg1}.

Our algorithm is referred to as the nonuniform proximal-averaged thresholding algorithm ({\it Nu}PATA-$1$), which relies on Majorization-Minimization (MM) \cite{figueiredo2007majorization} and the proximal average strategy \cite{bauschke2008proximal, yu2013better, jawali2020cornet}. Further, we unfold the {\it Nu}PATA-$1$ iterations into a learnable network called nonuniform sparse proximal average network ({\it Nu}SPAN-$1$).

\subsection{Optimization Algorithm -- Type-\texorpdfstring{$1$}{1}}
\label{subsec:nupata1}

Due to P1, there exists $\eta < 1/L$ such that $f(\boldsymbol{x})$ is upper bounded locally by a quadratic expansion about $\boldsymbol{x} = \boldsymbol{x}^{(k)}$ as follows:
\begin{equation}
f(\boldsymbol{x}) \leq Q\left(\boldsymbol{x},\boldsymbol{x}^{(k)}\right),
\end{equation}
where $Q(\boldsymbol{x},\boldsymbol{x}^{(k)}) = f(\boldsymbol{x}^{(k)}) + \langle \nabla f(\boldsymbol{x}^{(k)}) , \boldsymbol{x}-\boldsymbol{x}^{k} \rangle + \frac{1}{2\eta} \Vert\boldsymbol{x}- \boldsymbol{x}^{(k)}\Vert_{2}^{2}$. The majorizer to $\mathcal{J}$ at $\boldsymbol{x}^{(k)}$ is 
\begin{equation}
\mathcal{J}(\boldsymbol{x}) \leq \underbrace{Q(\boldsymbol{x},\boldsymbol{x}^{(k)}) + g(\boldsymbol{x})}_{H(\boldsymbol{x},\boldsymbol{x}^{(k)})}.
\end{equation}

The update for $\boldsymbol{x}$ at the $(k+1)^{\text{th}}$ iteration is given by 
\begin{align*}
\boldsymbol{x}^{(k+1)} & = \argminA_{\boldsymbol{x}} H(\boldsymbol{x},\boldsymbol{x}^{(k)}), \\
& = \argminA_{\boldsymbol{x}} \frac{1}{2 \eta} \left\|\boldsymbol{x} - \left( \boldsymbol{x}^{(k)}- \eta\nabla_{\boldsymbol{x}} f(\boldsymbol{x}^{(k)}) \right)\right\|_{2}^{2} \\ & \hspace{3cm} + g(\boldsymbol{x}). \numberthis \label{exactcompprob1}
\end{align*}
The above problem does not have a closed-form solution. Hence, we consider an approximate variant of the problem corresponding to Eq.~\eqref{exactcompprob1} based on the proximal average strategy:
\begin{align*}
\boldsymbol{x}^{(k+1)} 
 & = \arg\min_{\boldsymbol{x}} \sum\limits_{i = 1}^m \frac{\omega_{i}}{2 \eta} {\left\| \boldsymbol{x} - \left( \boldsymbol{x}^{(k)}- \eta\nabla_{\boldsymbol{x}} f(\boldsymbol{x}^{(k)}) \right) \right\|}_{2}^{2} \\
  & \hspace{2cm} + \sum\limits_{i = 1}^m { \omega_{i} g_{i}(\boldsymbol{x}) },\\
 & = \sum\limits_{i=1}^m \omega_i \mathcal{P}_{g_i}(\mathbf{u}^{(k)}) , \numberthis \label{nupata1}
\end{align*}
where $\mathbf{u}^{(k)} = \boldsymbol{x}^{(k)}- \eta\nabla_{\boldsymbol{x}} f(\boldsymbol{x}^{(k)})$, $ \nabla_{\boldsymbol{x}} f(\boldsymbol{x}^{(k)}) = - \boldsymbol{h}' * \left( \boldsymbol{y} - \boldsymbol{h} * \boldsymbol{x}^{(k)} \right)$, where $\boldsymbol{h}'$ is the flipped version of $\boldsymbol{h}$, and $\mathcal{P}_{g_i}$ represent the proximal operators corresponding to $g_i$ (cf. Table~\ref{table:regularizers}). The optimization procedure is listed in Algorithm~\ref{alg:nupata1}.

\begin{algorithm}[t]
    \caption{Nonuniform Sparse Proximal-Averaged Thresholding Algorithm -- Type-$1$ ({\it Nu}PATA-$1$)}\label{alg:nupata1}
    \KwIn{$\boldsymbol{y}$, $\boldsymbol{h}$, $L = {\left\| \boldsymbol{h} \right\|}_2^2, k_{max}$, $\{\omega_{i}: 0 < \omega_{i} < 1, \sum_{i=1}^m \omega_{i} = 1\}$, $\{\boldsymbol{\lambda$, $\mu$, $\nu}\} \in \mathbb{R}^n_{>0}$, $\{\boldsymbol{\gamma}\in \mathbb{R}^n_{>1}, \boldsymbol{a} \in \mathbb{R}^n_{>2} \}$}
    Initialize: $ \boldsymbol{x}^{(0)} = \boldsymbol{0} $\;
    \While{$k < k_{max}$}{
        $\boldsymbol{z}^{(k+1)} = \boldsymbol{x}^{(k)} + (1/2L) \left( \boldsymbol{h}'* (\boldsymbol{y} - \boldsymbol{h}*{\boldsymbol{x}^{(k)}}) \right)$\;
        $\boldsymbol{x}^{(k+1)} = \sum\limits_{i = 1}^m \omega_{i} \mathcal{P}_{g_i} \left( \boldsymbol{z}^{(k+1)} \right)$\;
        $k = k + 1$\;
        }
    \KwOut{$\hat{\boldsymbol{x}} \leftarrow \boldsymbol{x}^{(k_{max})}$}
\end{algorithm}%

\begin{algorithm}[t]
    \caption{Nonuniform Sparse Proximal Average Network -- Type-$1$ ({\it Nu}SPAN-$1$)}\label{alg:nuspan1}
    \KwIn{$\boldsymbol{y}$, $L = {\left\| \mathbf{H} \right\|}_2^2, k_{max}$, $\{\omega_{i}: 0 < \omega_{i} < 1, \sum_{i=1}^m \omega_{i} = 1\}$, $\{\boldsymbol{\lambda$, $\mu$, $\nu}\} \in \mathbb{R}^n_{>0}$, $\{\boldsymbol{\gamma}\in \mathbb{R}^n_{>1}, \boldsymbol{a} \in \mathbb{R}^n_{>2} \}$}
    Initialize: $\mathbf{W}, \mathbf{S}, \boldsymbol{x}^{(0)} = \sum\limits_{i = 1}^m \omega_{i} \mathcal{P}_{g_{i}} (\mathbf{W}{\boldsymbol{y}}) $\;
    \While{$k < k_{max}$}{
        $\mathbf{c}^{(k+1)} = \mathbf{W}{\boldsymbol{y}} + \mathbf{S}{\boldsymbol{x}^{(k)}} $\;
        $\boldsymbol{x}^{(k+1)} = \sum\limits_{i = 1}^m \omega_{i} \mathcal{P}_{g_{i}} (\mathbf{c}^{(k+1)}) $\;
        $k = k + 1$\;
        }
    $\hat{\boldsymbol{x}}_{\mathcal{S}(\hat{\boldsymbol{x}})} = \mathbf{H}_{\mathcal{S}(\hat{\boldsymbol{x}})}^{\dagger}\boldsymbol{y}$\;
    \KwOut{$\hat{\boldsymbol{x}} \leftarrow \boldsymbol{x}^{(k_{max})}$}
\end{algorithm}%

\subsection{Nonuniform Sparse Proximal Average Network -- Type-\texorpdfstring{$1$}{1} ({\it Nu}SPAN-\texorpdfstring{$1$}{1})}
\label{subsec:nu_span1}%
The update in \eqref{nupata1} involves convolutions followed by nonlinear activation, and can therefore be represented as a layer in a neural network. We unfold the iterations of Algorithm~\ref{alg:nupata1} to obtain the deep-unrolled architecture, namely, nonuniform sparse proximal average network ({\it Nu}SPAN-$1$) for solving the reflectivity inversion problem in Eq.~\eqref{nu_comp_reg1}. The structure for each layer in {\it Nu}SPAN-$1$ is given by
\begin{equation}
    \boldsymbol{x}^{(k+1)} = \sum_{i = 1}^m \omega_{i} \mathcal{P}_{g_{i}} ( \mathbf{W} \boldsymbol{y} + \mathbf{S} \boldsymbol{x}^{(k)}),
\end{equation}
where \( \mathbf{W}=(1/L)\mathop{\boldsymbol{h}'} \) and \( \mathbf{S}=\mathbf{I}-(1/L)\mathop{\boldsymbol{h}' * \boldsymbol{h}} \) \cite{gregor2010learning} are initialized as Toeplitz matrices. In the learning stage, they are dense and unstructured. Given training data that consists of a large number of independent and identically distributed examples $\{\boldsymbol{x}_p, \boldsymbol{y}_p\}_{p=1}^{N}$ and {\it Nu}SPAN-$1$ with a fixed number of layers, we optimize the smooth $\ell_1$ cost computed between the true reflectivity $\boldsymbol{x}$ and the prediction $\hat{\boldsymbol{x}}(\boldsymbol{\theta})$:
\begin{align*}
    \min_{\boldsymbol{\theta}}\frac{1}{N}\sum\limits_{i=1}^N \left(\beta \left\|{\boldsymbol{x} - \hat{\boldsymbol{x}}(\boldsymbol{\theta})} \right\|_{1} + (1-\beta) \left\|{\boldsymbol{x} - \hat{\boldsymbol{x}}(\boldsymbol{\theta})} \right\|_{2}^{2}\right),
\end{align*}
where $0 < \beta < 1$. We set $\beta = 1$; it is not a trainable parameter. The parameters $\boldsymbol{\theta}$ that need to be learned are the matrices $\mathbf{W}$ and $\mathbf{S}$, the parameters of $g_{i}$: $\{\boldsymbol{\lambda$, $\mu$, $\nu}\} \in \mathbb{R}^n_{>0}$ and $\{\boldsymbol{\gamma}\in \mathbb{R}^n_{>1}, \boldsymbol{a} \in \mathbb{R}^n_{>2} \}$, and the weights $\{\omega_{i}: 0 < \omega_{i} < 1, \sum_{i=1}^m \omega_{i} = 1\}$. The {\it Nu}SPAN-$1$ algorithm for $k_{max}$ layers is listed in Algorithm~\ref{alg:nuspan1}.%

We enhance the amplitude recovery of {\it Nu}SPAN-$1$ (and {\it Nu}SPAN-$2$ in Section~\ref{sec:nu_span2}) during the inference phase by re-estimating the amplitudes over the supports given by {\it Nu}SPAN-$1$ as: $\hat{\boldsymbol{x}}_{\mathcal{S}(\hat{\boldsymbol{x}})} = \mathbf{H}_{\mathcal{S}(\hat{\boldsymbol{x}})}^{\dagger}\boldsymbol{y}$, where $\hat{\boldsymbol{x}}$ is the sparse vector estimated by {\it Nu}SPAN-$1$, $\mathcal{S}(\hat{\boldsymbol{x}})$ is the support, i.e., non zero locations of $\hat{\boldsymbol{x}}$ , and $\mathbf{H}_{\mathcal{S}(\hat{\boldsymbol{x}})}^{\dagger}$ is the pseudo-inverse of the Toeplitz matrix $\mathbf{H}$ of the kernel $\boldsymbol{h}$ over $\mathcal{S}(\hat{\boldsymbol{x}})$.

\begin{figure*}[t]
    \centering
    \resizebox{\linewidth}{!}{
    \includegraphics{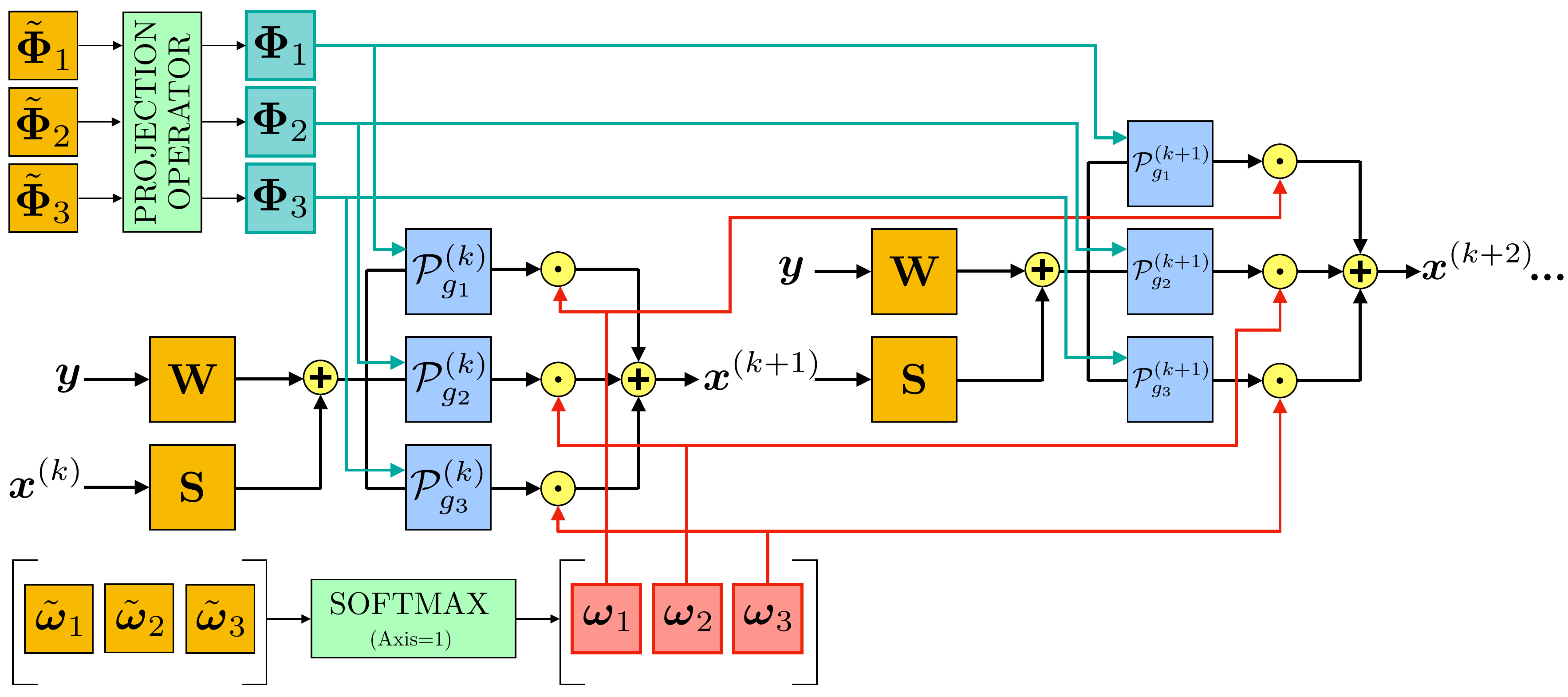}}
    \caption{\protect\rgbsymbol~Structure of layers in {\it Nu}SPAN-$2$. The matrices $\mathbf{W}$, $\mathbf{S}$, and $\left[ \boldsymbol{\omega}_1 ~ \boldsymbol{\omega}_2 ~ \boldsymbol{\omega}_3 \right]$, and $\boldsymbol{\Phi}_i$, which are the parameters of $g_i$: $\{\boldsymbol{\lambda$, $\mu$, $\nu}\} \in \mathbb{R}^n_{>0}$ and $\{\boldsymbol{\gamma}\in \mathbb{R}^n_{>1}, \boldsymbol{a} \in \mathbb{R}^n_{>2} \}$ (corresponding to the penalties in Table~\ref{table:regularizers}), are shared across layers of {\it Nu}SPAN-$2$. The projection operator refers to the constraints imposed according to Table~\ref{table:regularizers} through the Pytorch \cite{paszke2019pytorch} clamp function. The operators $\oplus$ and $\odot$ represent sum and product, all considered element-wise.}
    \label{fig:schematic2}
\end{figure*}

\begin{algorithm}[t]
    \caption{Nonuniform Sparse Proximal-Averaged Thresholding Algorithm -- Type-$2$ ({\it Nu}PATA-$2$)}\label{alg:nupata2}
    \KwIn{$\boldsymbol{y}$, $\boldsymbol{h}$, $L = {\left\| \boldsymbol{h} \right\|}_2^2, k_{max}$, $\{\boldsymbol{\omega}_{i}: \boldsymbol{0} < \omega_{i} < \boldsymbol{1}, \sum_{i=1}^m \boldsymbol{\omega}_{i} = \boldsymbol{1}\}$, $\{\boldsymbol{\lambda$, $\mu$, $\nu}\} \in \mathbb{R}^n_{>0}$, $\{\boldsymbol{\gamma}\in \mathbb{R}^n_{>1}, \boldsymbol{a} \in \mathbb{R}^n_{>2} \}$}
    Initialize: $ \boldsymbol{x}^{(0)} = \boldsymbol{0} $\;
    \While{$k < k_{max}$}{
        $\boldsymbol{z}^{(k+1)} = \boldsymbol{x}^{(k)} + (1/2L) \left( \boldsymbol{h}'* (\boldsymbol{y} - \boldsymbol{h}*{\boldsymbol{x}^{(k)}}) \right)$\;
        $\boldsymbol{x}^{(k+1)} = \sum\limits_{i = 1}^m \boldsymbol{\omega}_{i} \odot \mathcal{P}_{g_i} \left( \boldsymbol{z}^{(k+1)} \right)$\;
        $k = k + 1$\;
        }
    \KwOut{$\hat{\boldsymbol{x}} \leftarrow \boldsymbol{x}^{(k_{max})}$}
\end{algorithm}%

\begin{algorithm}[t]
    \caption{Nonuniform Sparse Proximal Average Network -- Type-$2$ ({\it Nu}SPAN-$2$)}\label{alg:nuspan2}
    \KwIn{$\boldsymbol{y}$, $L = {\left\| \mathbf{H} \right\|}_2^2, k_{max}$, $\{\boldsymbol{\omega}_{i}: \boldsymbol{0} < \omega_{i} < \boldsymbol{1}, \sum_{i=1}^m \boldsymbol{\omega}_{i} = \boldsymbol{1}\}$, $\{\boldsymbol{\lambda$, $\mu$, $\nu}\} \in \mathbb{R}^n_{>0}$, $\{\boldsymbol{\gamma}\in \mathbb{R}^n_{>1}, \boldsymbol{a} \in \mathbb{R}^n_{>2} \}$}
    Initialize: $\mathbf{W}, \mathbf{S}, \boldsymbol{x}^{(0)} = \sum\limits_{i = 1}^m \boldsymbol{\omega}_{i} \odot \mathcal{P}_{g_{i}} (\mathbf{W}{\boldsymbol{y}}) $\;
    \While{$k \leq k_{max}$}{
        $\mathbf{c}^{(k+1)} = \mathbf{W}{\boldsymbol{y}} + \mathbf{S}{\boldsymbol{x}^{(k)}} $\;
        $\boldsymbol{x}^{(k+1)} = \sum\limits_{i = 1}^m \boldsymbol{\omega}_{i} \odot \mathcal{P}_{g_{i}} (\mathbf{c}^{(k+1)}) $\;
        $k = k + 1$\;
        }
    $\hat{\boldsymbol{x}}_{\mathcal{S}(\hat{\boldsymbol{x}})} = \mathbf{H}_{\mathcal{S}(\hat{\boldsymbol{x}})}^{\dagger}\boldsymbol{y}$\;
    \KwOut{$\hat{\boldsymbol{x}} \leftarrow \boldsymbol{x}^{(k_{max})}$}
\end{algorithm}%

\section{Proximal Average Network for Reflectivity Inversion --- Nonuniform Sparse Model (Type-\texorpdfstring{$2$}{2})}
\label{sec:nu_span2}
The generalized variant of the composite-regularization problem given in Eq.~\eqref{nu_comp_reg1} is given as follows:
\begin{align*}
    \arg\min_{\boldsymbol{x}, \,\{\boldsymbol{\omega}_{i}\}} & \, {\frac{1}{2}{\left\| \boldsymbol{h}*\boldsymbol{x}-\boldsymbol{y} \right\|}_{2}^{2}} + \sum\limits_{i = 1}^m \langle \boldsymbol{\omega}_i, q_i(\boldsymbol{x}) \rangle \\
    & \text{s.t.} \sum\limits_{i=1}^m {\boldsymbol{\omega}_{i}=\boldsymbol{1}}, \boldsymbol{0}<\boldsymbol{\omega}_{i}<\boldsymbol{1}, \; \forall{i}, \numberthis \label{nu_comp_reg2}
\end{align*}%
where $q_i(\boldsymbol{x}) = [g_i(x_1), g_i(x_2), \ldots , g_i(x_n)]^{\textsc{T}}$, $g_i$ is the scalar penalty provided in Table~\ref{table:regularizers}, $\langle \cdot, \cdot \rangle$ denotes inner-product, $\boldsymbol{0}$ is the null vector, $\boldsymbol{1}$ is the vector of all ones, and $\boldsymbol{\omega}_i = [\omega_{i1}, \omega_{i2}, \ldots , \omega_{in}]^{\textsc{T}}$.

\subsection{Optimization algorithm -- Type-\texorpdfstring{$2$}{2}}
\label{subsec:nupata2}

We can follow the strategy mentioned in Section~\ref{subsec:nupata1} for optimizing the cost given in Eq.~\eqref{nu_comp_reg2}.
The update for $\boldsymbol{x}$ at $(k+1)^{\text{th}}$ iteration is given by 
\begin{align*}
\boldsymbol{x}^{(k+1)} & = \argminA_{\boldsymbol{x}} H \left( \boldsymbol{x},\boldsymbol{x}^{(k)} \right), \\
 & = \argminA_{\boldsymbol{x}} \frac{1}{2 \eta} \left\|\boldsymbol{x} - \left( \boldsymbol{x}^{(k)}- \eta\nabla_{\boldsymbol{x}} f(\boldsymbol{x}^{(k)}) \right) \right\|_{2}^{2} \\ & \hspace{3cm} + \sum\limits_{i = 1}^m \langle \boldsymbol{\omega}_i, q_i(\boldsymbol{x})\rangle. \numberthis \label{exactcompprob2}
\end{align*}
Since the above problem is complex and non-convex, we solve the approximate variant of the problem corresponding to Eq.~\eqref{exactcompprob2} based on the proximal average strategy \cite{yu2013better}.
\begin{align*}
\boldsymbol{x}^{(k+1)} 
 & = \arg\min_{\boldsymbol{x}} \sum\limits_{i = 1}^m \frac{\omega_{i}}{2 \eta} {\left\| \boldsymbol{x} -\left( \boldsymbol{x}^{(k)}- \eta\nabla_{\boldsymbol{x}} f(\boldsymbol{x}^{(k)}) \right) \right\|}_{2}^{2} \\
 & \hspace{2cm} + \sum\limits_{i = 1}^m \langle \boldsymbol{\omega}_i, q_i(\boldsymbol{x})\rangle,\\
 & = \sum\limits_{i=1}^m \boldsymbol{\omega}_i \odot \mathcal{P}_{g_i}(\mathbf{u}^{(k)}) . \numberthis \label{nupata2}
\end{align*}
Algorithm \eqref{alg:nupata2} illustrates the {\it Nu}PATA-$2$ formulation for solving the optimization problem \eqref{nu_comp_reg2} using the proximal average strategy. As in the previous section, we unfold the update in \eqref{nupata2} into a layer of a neural network.

\subsection{Nonuniform Sparse Proximal Average Network -- Type-\texorpdfstring{$2$}{2} ({\it Nu}SPAN-\texorpdfstring{$2$}{2})}
\label{subsec:nu_span2}%
Representing the {\it Nu}PATA-$2$ update in \eqref{nupata2} as a layer in a neural network, we obtain the nonuniform sparse proximal average network ({\it Nu}SPAN-$2$). The structure for each layer in {\it Nu}SPAN-$2$ is given by
\begin{equation}
    \boldsymbol{x}^{(k+1)} = \sum_{i = 1}^m \boldsymbol{\omega}_{i} \odot \mathcal{P}_{g_{i}} ( \mathbf{W} \boldsymbol{y} + \mathbf{S} \boldsymbol{x}^{(k)}),
    % \label{ista}
\end{equation}
where \( \mathbf{W} \) and \( \mathbf{S} \) definitions are provided in Section~\ref{subsec:nu_span1}. Similar to {\it Nu}SPAN-$1$, we optimize the learnable parameters of {\it Nu}SPAN-$2$ subject to the constraints given in Eq.~\eqref{nu_comp_reg2}, by minimizing the smooth $\ell_1$ cost defined between the true reflectivity $\boldsymbol{x}$ and prediction $\hat{\boldsymbol{x}}(\boldsymbol{\theta})$, where $\boldsymbol{\theta}$ denotes the learnable parameters such as the matrices $\mathbf{W}$ and $\mathbf{S}$, the parameters of $g_{i}(\cdot) \, \forall i$: $\{\boldsymbol{\lambda$, $\mu$, $\nu}\} \in \mathbb{R}^n_{>0}$ and $\{\boldsymbol{\gamma}\in \mathbb{R}^n_{>1}, \boldsymbol{a} \in \mathbb{R}^n_{>2} \}$, and the weights $\{\boldsymbol{\omega}_{i}: \boldsymbol{0} < \omega_{i} < \boldsymbol{1}, \sum_{i=1}^m \boldsymbol{\omega}_{i} = \boldsymbol{1}\}$. Algorithm~\ref{alg:nuspan2} gives the {\it Nu}SPAN-$2$ algorithm for $k_{max}$ layers, and Figure~\ref{fig:schematic2} gives the structure of layers in {\it Nu}SPAN-$2$. 

\section{Experimental Results}
\label{sec:results}
We demonstrate the efficacy of the proposed networks, namely, {\it Nu}SPAN-$1$ and {\it Nu}SPAN-$2$ on both synthetic and simulated datasets as well as on real data in comparison with the benchmark techniques such as BPI \cite{chen2001atomic, zhang2011seismic}, FISTA \cite{beck2009fast, perez2012inversion}, and SBL-EM \cite{wipf2004sparse, yuan2019seismic}. The performance of the proposed approaches is quantified based on objective metrics computed between ground-truth sparse vector $\boldsymbol{x}$ and the predicted sparse vector $\hat{\boldsymbol{x}}$ listed in the following section.

\subsection{Objective Metrics}
\label{metrics}
We evaluate the performance in terms of amplitude and support recovery metrics, which are crucial to estimate the amplitudes and locations of reflection coefficients.

\begin{itemize}[wide, labelwidth=0pt, labelindent=0pt, noitemsep, nolistsep]
    \item {Correlation Coefficient (CC) \cite{freedman2007statistics}:
    \begin{align*}
        \text{CC} = \frac{\langle {\boldsymbol{x}}, \hat{{\boldsymbol{x}}} \rangle - \langle {\boldsymbol{x}}, {{\boldsymbol{1}}} \rangle \langle \hat{{\boldsymbol{x}}}, {{\boldsymbol{1}}} \rangle}{\sqrt{\left( \left\| {{\boldsymbol{x}}} \right\|_2^2 - {\langle {\boldsymbol{x}}, {{\boldsymbol{1}}} \rangle}^2 \right) \left( \left\| \hat{{\boldsymbol{x}}} \right\|_2^2 - {\langle \hat{{\boldsymbol{x}}}, {{\boldsymbol{1}}} \rangle}^2 \right) }},
    \end{align*}%    
    }
    \item Relative Reconstruction Error (RRE) and Signal-to-Reconstruction Error Ratio (SRER) (in dB)%
    \begin{align*}%\label{rel_error}
        \text{RRE} = \frac{{\left\| \hat{\boldsymbol{x}} - \boldsymbol{x} \right\|}_2^2}{{\left\| \boldsymbol{x} \right\|}_2^2}, \text{SRER} = 10\mathop {\log\limits_{10} \left(\frac{{\left\| {\boldsymbol{x}} \right\|_2^2}}{{\left\| {\hat{\boldsymbol{x}} - {\boldsymbol{x}}} \right\|_2^2}}\right)}.
    \end{align*}
    \item {Probability of Error in Support (PES) is given by%
    \begin{align*}
        \text{PES} = \frac{1}{t} \mathop{ \sum\limits_{i = 1}^t \left( \frac{{{\text{max}}(\left| {\mathcal{S}({\hat{\boldsymbol{x}_i}})} \right|,\left| {\mathcal{S}({\boldsymbol{x}_i})} \right|) - \left| {\mathcal{S}({\hat{\boldsymbol{x}_i}})\mathop {\cap} {\mathcal{S}}({\boldsymbol{x}_i})} \right|}}{{{\text{max}}(\left| {\mathcal{S}({\hat{\boldsymbol{x}_i}})} \right|,\left| {\mathcal{S}({\boldsymbol{x}_i})} \right|)}}\right)}, 
        % \label{pes}
    \end{align*}%
    where $|\cdot|$ denotes the cardinality, and $\mathcal{S}(\cdot)$ denotes the support of the argument.}
\end{itemize}%

\begin{figure*}[!ht]
    \centering
    \resizebox{0.861\linewidth}{!}{
    \includegraphics{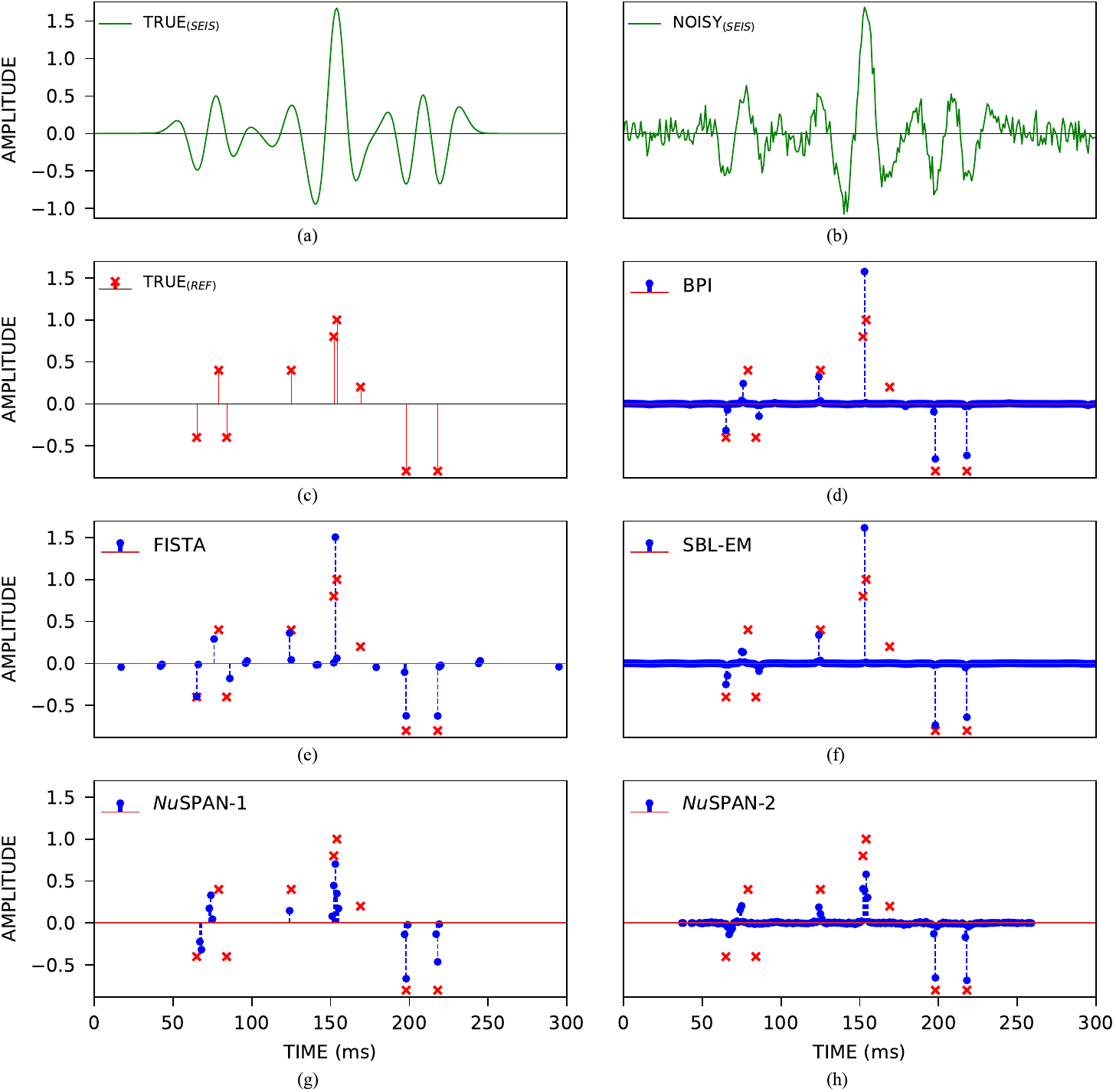}}%CCs: BPI[0.3261], FISTA[0.3513], SBL-EM[0.3336], {\it Nu}SPAN-$1$[0.6226], {\it Nu}SPAN-$2$[0.7856]
    \caption{\protect\rgbsymbol~Sample results for a synthetic $1$-D seismic trace. (a) True seismic trace; (b) noisy seismic trace; (c) true reflectivity; (d)-(h) Recovered reflectivity (blue circles) compared with true reflectivity (red crosses). The proposed techniques, {\it Nu}SPAN-$1$ (g) and {\it Nu}SPAN-$2$ (h), distinguish between closely-spaced spikes around $150$ ms; the benchmark techniques BPI, FISTA, and SBL-EM (d)-(f) predict a single reflector instead.}
    \label{fig:trace}
\end{figure*}

\subsection{Training Phase}
The synthetic training data that appropriately represents observed seismic data is generated as recommended by \cite{kim2018geophysical}. We generate synthetic training data of size $5 \times 10^{5}$ seismic traces, each consisting of $300$ samples, obtained by convolving $1$-D reflectivity profiles with a Ricker wavelet. Each reflectivity profile contains $200$ samples with amplitudes ranging from $-1.0$ to $1.0$, padded with zeros before convolution with the wavelet to have the same length as the seismic trace \cite{russell2019machine}. The sparsity factor, which is the ratio of the number of non-zero elements to the total number of elements, is set to $0.05$. We also report results for sparsity factors $0.10$, $0.15$, and $0.20$ in the supplementary document. The reflector/spike locations are chosen uniformly at random, without any minimum spacing constraint between any two spikes. Reflectivity values are then assigned to these locations, picked randomly from the amplitude range of $-1.0$ to $1.0$. The amplitude increment, sampling interval, wavelet frequency, and the initial hyperparameters vary depending on the dataset. The optimum number of layers also varies with the dataset, and we consider $10$, $15$, or $20$ layer-models in our experiments. Our models are trained with a batch size of $200$, use the ADAM optimizer \cite{kingma2014adam}, and with the learning rate set to $1 \times 10^{-3}$, and input measurement SNR set to $10$ dB --- to ensure robustness against noisy testing data \cite{kim2018geophysical}. We consider models trained on $1$-D data to operate trace-by-trace on all the datasets: synthetic $1$-D data, synthetic $2$-D wedge models \cite{hamlyn2014thin}, simulated Marmousi$2$ data \cite{martin2006marmousi2}, and real data.

\subsection{Testing Phase -- Synthetic \texorpdfstring{$1$}{1}-D \& \texorpdfstring{$2$}{2}-D Data}
\label{results:synthetic}

\subsubsection{Synthetic \texorpdfstring{$1$}{1}-D Traces}
\label{subsubsec:results_1d}
We validate the performance of the proposed models over $1000$ realizations of synthetic $1$-D traces, with a $30$ Hz Ricker wavelet, $1$ ms sampling interval, $0.2$ amplitude increments, and the minimum spacing between reflection coefficients (spikes) $1$ ms. Comparisons across several methods are reported in Figure~\ref{fig:trace} and Table~\ref{table:1d_compare}. Figure~\ref{fig:trace} shows the results for a sample synthetic $1$-D trace out of the $1000$ test realizations for which the results are reported in Table~\ref{table:1d_compare}. {\it Nu}SPAN-$1$ and {\it Nu}SPAN-$2$ resolve closely-spaced reflection coefficients, whereas the benchmark techniques (BPI, FISTA, and SBL-EM) predict a single reflector. From Table~\ref{table:1d_compare}, observe that {\it Nu}SPAN-$1$ and {\it Nu}SPAN-$2$ (Nonuniform Sparse Proximal Average Network, Type $1$ and $2$) recover amplitudes with higher accuracy than the benchmark methods, and {\it Nu}SPAN-$1$ exhibits superior support recovery. The low computation time of the proposed models is crucial for processing the large volume of data in reflection seismic processing.

\begin{table}[t]
    \centering
    \caption{Performance evaluation in terms of objective metrics averaged over $1000$ test realizations of synthetic $1$-D traces. The proposed {\it Nu}SPAN-$1$ and {\it Nu}SPAN-$2$ show superior amplitude recovery in terms of CC, RRE, and SRER and significantly lower computation time, while {\it Nu}SPAN-$1$ outperforms in support recovery in terms of PES. The best performance is highlighted in \textbf{boldface}. The second best is \underline{underlined}.}
    \label{table:1d_compare}
    \resizebox{\columnwidth}{!}{
    \begin{tabular}{l|c|c|c|c|c}
        \toprule
        \bfseries Method & \bfseries CC & \bfseries RRE & \bfseries SRER & \bfseries PES & \bfseries Time ($\mathrm{s}$) \\
        \midrule
        BPI & ${0.5499}$ & ${0.7290}$ & ${1.8687}$ & ${0.9704}$ & ${503.8519}$\\
        FISTA & ${0.5473}$ & ${0.7203}$ & ${1.8391}$ & $\underline{0.8112}$ & ${33.2539}$\\
        SBL-EM & ${0.5501}$ & ${0.7682}$ & ${1.8276}$ & ${0.9704}$ & ${1076.1833}$\\
        {\it Nu}SPAN-$1$ & $\underline{0.5979}$ & $\underline{0.6354}$ & $\underline{2.2038}$ & $\mathbf{0.7104}$ & $\mathbf{0.1778}$\\%10-layer model
        {\it Nu}SPAN-$2$ & $\mathbf{0.6050}$ & $\mathbf{0.6274}$ & $\mathbf{2.2508}$ & ${0.9563}$ & $\underline{0.1870}$\\%10-layer model
        \bottomrule
    \end{tabular}}
\end{table}

\subsubsection{Synthetic \texorpdfstring{$2$}{2}-D Wedge Models}
\label{subsubsec:results_wedge}
Synthetic $2$-D wedge models are considered to evaluate resolving capability on thin beds \cite{hamlyn2014thin}. A wedge model typically consists of two interfaces, one horizontal and another inclined, with the polarity (N:~Negative, P:~Positive) of reflection coefficients same or opposite, giving four possible types of wedge models. Here, we report results from an odd wedge model (NP) with negative polarity on the upper horizontal interface (N) and positive on the lower inclined interface (P), and an even (NN) wedge model, while those for the remaining two variants are given in the supplementary. In our experimental setup, each model consists of $26$ traces, with separation between reflectors of the two interfaces increasing from $0$ ms to $50$ ms in $2$ ms increments. The amplitudes of the reflectors are fixed as $\pm~0.5$ based on the polarity. 

\begin{figure*}[t]
    \centering
    \resizebox{0.6\linewidth}{!}{
    \includegraphics[width=\linewidth]{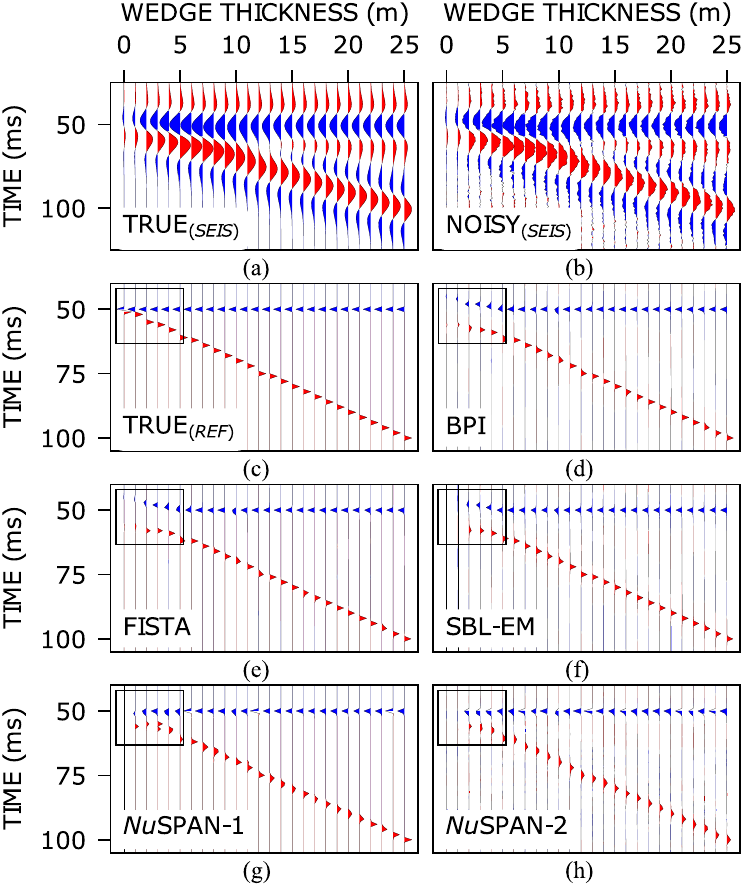}}
    \caption{\protect\rgbsymbol~Results for a synthetic $2$-D odd (NP) wedge model. True (a) seismic traces and (c) reflectivity. (b) Noisy seismic traces with input SNR $10$ dB. (d)-(h) Recovered reflectivity profiles show that {\it Nu}SPAN-$1$ (g) and {\it Nu}SPAN-$2$ (h) resolve reflectors $< 5$ m thickness, whereas the baselines (d)-(f) fail to do so, evident from the diverging interfaces highlighted by the rectangle in black.}
    \label{fig:wedge_np}
\end{figure*}

Results for the NP odd wedge model are given in Table~\ref{table:np} and Figure~\ref{fig:wedge_np}. {\it Nu}SPAN-$1$ outperforms other benchmark techniques in terms of the objective metrics, namely, CC and RRE, and shows superior support recovery, measured in terms of PES. Figure~\ref{fig:wedge_np} highlights the fact that BPI, FISTA, and SBL-EM fail to resolve the locations of closely-spaced reflectors, which is evident from the divergence observed below the tuning thickness of $13$ ms (wedge thickness between $5$-$6$ m) \cite{chung1995frequency}.

\begin{figure*}[t]
    \centering
    \resizebox{.6\linewidth}{!}{
    \includegraphics{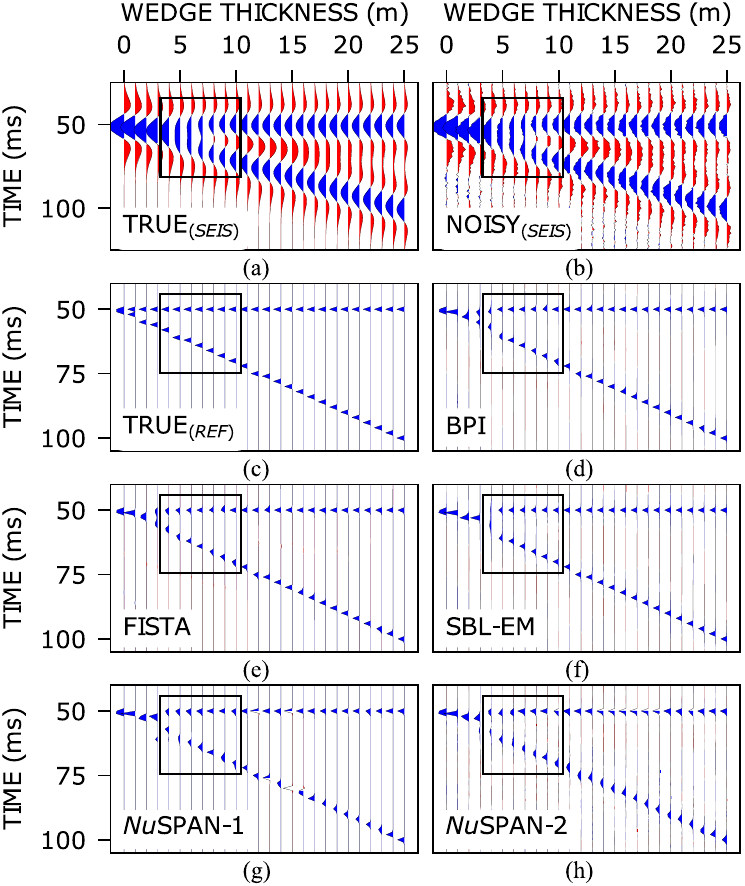}}
    \caption{\protect\rgbsymbol~Results for a synthetic $2$-D even (NN) wedge model. True (a) seismic traces and (c) reflectivity; (b) Noisy seismic traces with input SNR $10$ dB; (d)-(h) Recovered reflectivity. The underperformance of {\it Nu}SPAN-$1$ and {\it Nu}SPAN-$2$ reported in Table~\ref{table:nn} could be due to the poor amplitude recovery, as a result of the destructive interference between the two interfaces, in the region highlighted in black.}
    \label{fig:wedge_nn}
\end{figure*}

Table~\ref{table:nn} and Figure~\ref{fig:wedge_nn} give the results for the NN even wedge model. For the even wedge model, the baselines perform marginally better than {\it Nu}SPAN-$1$ in terms of amplitude recovery, measured through CC, RRE and SRER, but {\it Nu}SPAN-$1$ outperforms in terms of support recovery (Table~\ref{table:nn}). Figure~\ref{fig:wedge_nn} shows that the baselines and the proposed networks recover the reflectivity profile well for the even wedge model. 

Although {\it Nu}SPAN-$1$ outperforms the baselines in all the objective metrics except SRER, we observe a drop in the performance of {\it Nu}SPAN-$1$ and {\it Nu}SPAN-$2$ in the synthetic $2$-D case over the $1$-D case in Section~\ref{subsubsec:results_1d}, especially in terms of SRER. We hypothesize that this could be attributed to the mismatch between the training and testing conditions. Further, we observe a disparity between the performance of the proposed networks in the case of the odd (NP) vs. even (NN) wedge model. This discrepancy could be attributed to the suboptimal amplitude recovery of reflection coefficients in the region where destructive interference is observed between the two interfaces of the wedge models (as highilghted in Figure~\ref{fig:wedge_nn}). Nonetheless, the support recovery evaluated in terms of PES remains comparable in the case of both wedge models (compare Tables~\ref{table:np} and \ref{table:nn}). However, these aspects need further investigation. 

\begin{table}[t]
    \centering
    \caption{Metrics for a synthetic $2$-D odd (NP) wedge model. {\it Nu}SPAN-$1$ outperforms in amplitude recovery in terms of CC and RRE and support recovery in terms of PES. The best performance is highlighted in \textbf{boldface}. The second best is \underline{underlined}.}
    \label{table:np}
    \resizebox{0.85\columnwidth}{!}{
    \begin{tabular}{l|c|c|c|c}
        \toprule
        \bfseries Method & \bfseries CC & \bfseries RRE & \bfseries SRER & \bfseries PES\\
        \midrule
        BPI & ${0.7865}$ & ${0.2919}$ & $\mathbf{12.3197}$ & ${0.9933}$\\
        FISTA & ${0.7786}$ & ${0.3008}$ & $\underline{12.0683}$ & $\underline{0.7090}$\\
        SBL-EM & $\underline{0.8064}$ & $\underline{0.2693}$ & ${10.1708}$ & ${0.9933}$\\
        {\it Nu}SPAN-$1$ & $\mathbf{0.8736}$ & $\mathbf{0.2553}$ & ${8.4383}$ & $\mathbf{0.5994}$\\%15-layer model
        {\it Nu}SPAN-$2$ & ${0.7858}$ & ${0.3441}$ & ${5.3164}$ & ${0.9896}$\\%10-layer model
        \bottomrule
    \end{tabular}}
\end{table}

\subsection{Testing Phase -- Marmousi\texorpdfstring{$2$}{2} Model}
\label{results:marmousi2}

The Marmousi$2$ model \cite{martin2006marmousi2} is widely used in reflection seismology to calibrate algorithms in structurally complex settings. The model (width $\times$ depth: $17$ km $\times~3.5$ km), an expanded version of the original Marmousi model (width $\times$ depth: $9.2$ km $\times~3$ km) \cite{versteeg1994marmousi}, has a $2$ ms sampling interval, with traces at an interval of $12.5$ m. We obtained the reflectivity profile from the P-wave velocity and density models \eqref{rhov} and convolved them with a $30$ Hz Ricker wavelet to generate the measurement. 

\begin{figure*}[t]
    \centering
    \resizebox{0.985\linewidth}{!}{
    \includegraphics{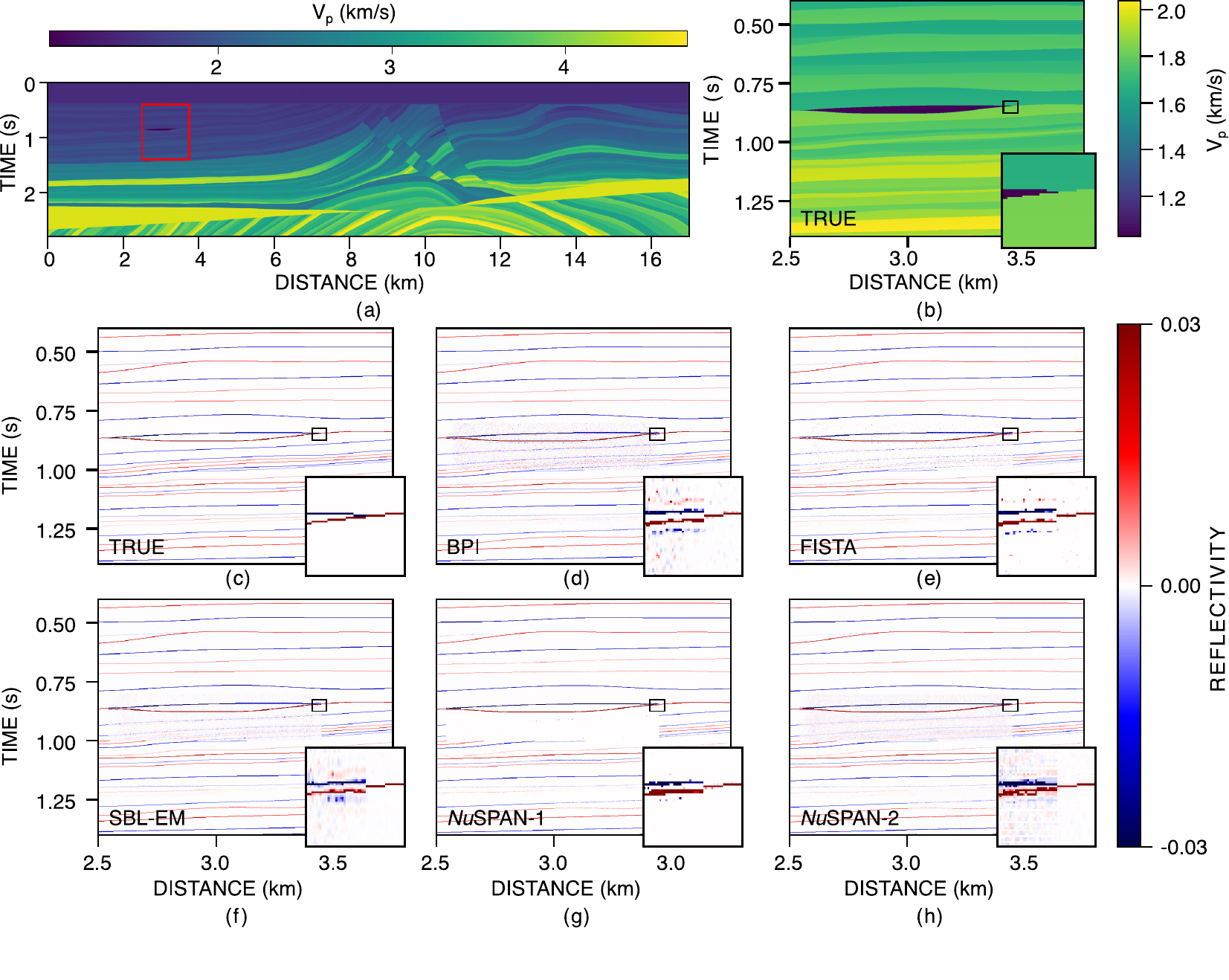}}
    \caption{\protect\rgbsymbol~P-wave velocity profile for the (a) Marmousi$2$ model and (b) the portion corresponding to Table~\ref{table:mm_mute}. (c) True reflectivity (ground truth), and (d)-(h) predicted reflectivity. The inset plots are zoomed-in portions of the selected area. {\it Nu}SPAN-$1$ and {\it Nu}SPAN-$2$ preserve lateral continuity better.}
    \label{fig:mm}
\end{figure*}

Figure~\ref{fig:mm} shows the result for a region of the model with a gas-charged sand channel \cite{martin2006marmousi2}. {\it Nu}SPAN-$1$ and {\it Nu}SPAN-$2$ preserve the lateral continuity better, evident from an observation of the insets in Figure~\ref{fig:mm}. BPI and FISTA introduce false interfaces due to the interference of multiple events at the ends of the channel. The testing times mentioned in Table~\ref{table:1d_compare} for synthetic data, when computed for the Marmousi$2$ model, further highlight the advantage of the proposed approach for seismic processing (Table~\ref{table:mm_mute}).

\begin{table}[t]
    \centering
    \caption{Results for a synthetic $2$-D odd (NN) wedge model. {\it Nu}SPAN-$1$ offers superior support recovery in terms of PES. The benchmark techniques (BPI, FISTA, and SBL-EM) show higher amplitude recovery accuracy quantified in terms of CC, RRE, and SRER.}
    \label{table:nn}
    \resizebox{0.85\columnwidth}{!}{
    \begin{tabular}{l|c|c|c|c}
        \toprule
        \bfseries Method & \bfseries CC & \bfseries RRE & \bfseries SRER & \bfseries PES\\
        \midrule
        BPI & ${0.8504}$ & $\underline{0.2831}$ & $\mathbf{12.4768}$ & ${0.9933}$\\
        FISTA & $\mathbf{0.8791}$ & $\mathbf{0.2524}$ & $\underline{11.7202}$ & $\underline{0.6681}$\\
        SBL-EM & $\underline{0.8519}$ & ${0.3280}$ & ${10.7475}$ & ${0.9933}$\\
        {\it Nu}SPAN-$1$ & ${0.8208}$ & ${0.3963}$ & ${9.0663}$ & $\mathbf{0.5885}$\\
        {\it Nu}SPAN-$2$ & ${0.7591}$ & ${0.4605}$ & ${5.5752}$ & ${0.9896}$\\
        \bottomrule
    \end{tabular}}
\end{table}

\begin{table}[t]
    \centering
    \caption{Metrics for a portion of the Marmousi$2$ model. {\it Nu}SPAN-$2$ outperforms in terms of CC, while {\it Nu}SPAN-$1$ shows superior support recovery in terms of PES. The reported time is for the complete model, showing significantly lower computation time for {\it Nu}SPAN-$1$ and {\it Nu}SPAN-$2$. The best performance is highlighted in \textbf{boldface}. The second best is \underline{underlined}.}%(mute)
    \label{table:mm_mute}
    \resizebox{\columnwidth}{!}{
    \begin{tabular}{l|c|c|c|c|c}
        \toprule
        \bfseries Method & \bfseries CC & \bfseries RRE & \bfseries SRER & \bfseries PES & \bfseries Time ($\mathrm{h}$)\\
        \midrule
        BPI & ${0.9473}$ & ${0.0875}$ & ${14.8024}$ & ${0.9724}$ & ${16.2671}$ \\% (58561.5337 s)
        FISTA & ${0.9407}$ & ${0.1017}$ & ${13.8579}$ & $\underline{0.7146}$ & ${5.5636}$ \\% (20028.9425 s)
        SBL-EM & $\underline{0.9549}$ & $\mathbf{0.0684}$ & $\mathbf{18.1890}$ & ${0.9724}$ & ${101.2659}$ \\% (364557.194 s)
        {\it Nu}SPAN-$1$ & ${0.9376}$ & ${0.1032}$ & ${14.2592}$ & $\mathbf{0.3693}$ & $\mathbf{0.0737}$\\% (265.3918 s)
        {\it Nu}SPAN-$2$ & $\mathbf{0.9591}$ & $\underline{0.0715}$ & $\underline{15.0973}$ & ${0.9662}$ & $\underline{0.1024}$\\% (368.8136 s)
        \bottomrule
    \end{tabular}}
\end{table}

\subsection{Testing Phase -- Real Data}
\label{results:real}
We validate on real data from the field, a $3$-D volume from the Penobscot $3$-D survey off the coast of Nova Scotia, Canada \cite{penobscot3d}. We pick a portion of the $3$-D volume, with $201$ Inlines (from inline $1150$-$1350$) and $121$ Xlines (between $1040$-$1160$), chosen such that the region includes two wells (wells L-$30$ and B-$41$) \cite{bianco2014geophysical}. The sampling interval for the dataset is $4$ ms, and the region chosen includes $800$ samples between $0$-$3196$ ms along the time/depth axis. A $25$ Hz Ricker wavelet fits the data well, also observed by \cite{bianco2014geophysical}.

\begin{figure*}[!ht]
    \centering
    \resizebox{0.985\linewidth}{!}{
    \includegraphics{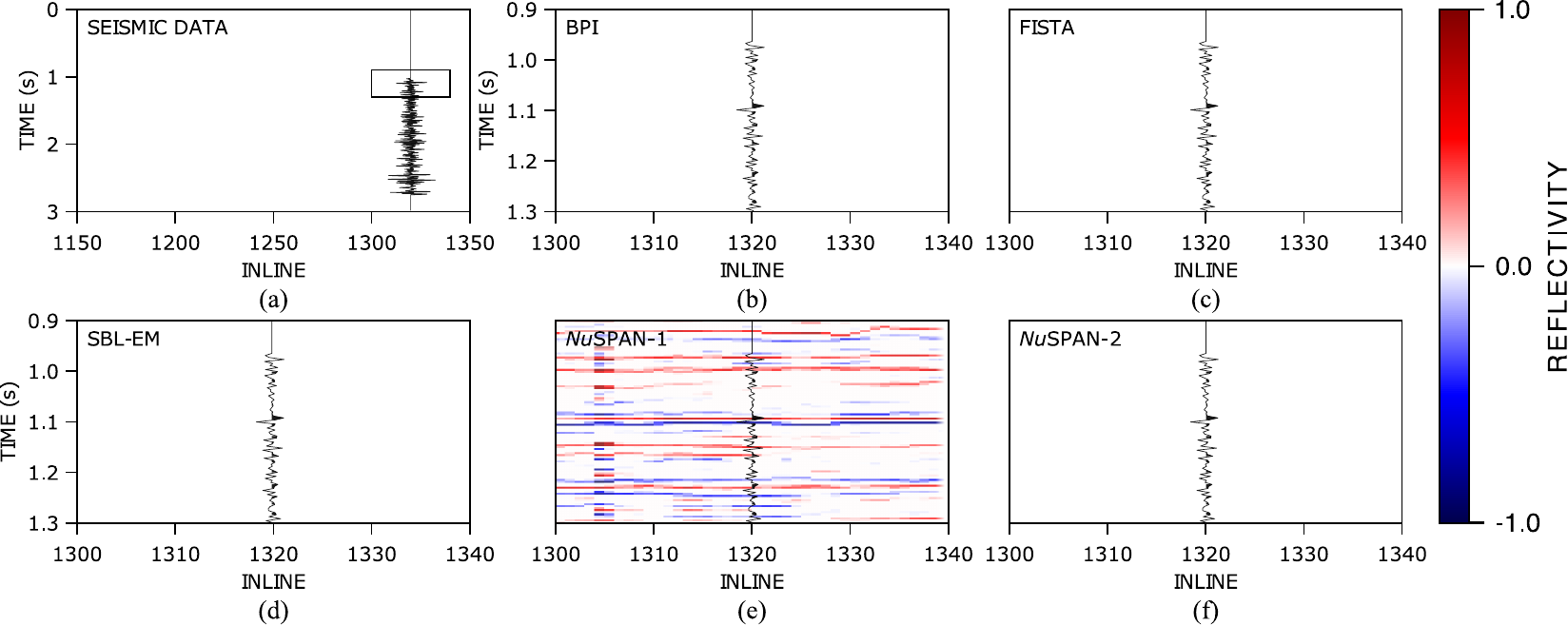}}
    \caption{\protect\rgbsymbol~(a) Observed seismic data, and (b)-(f) predicted reflectivity profiles for the inset marked in (a) for Xline $1155$ of the Penobscot $3$-D survey \cite{penobscot3d}. The overlaid waveforms in black show the recorded (a) well seismic and (b)-(f) reflectivity profiles, respectively, at well L-30 \cite{bianco2014geophysical}. {\it Nu}SPAN-$1$ and {\it Nu}SPAN-$2$ show superior amplitude recovery while also removing noise from the observations.} 
    \label{fig:real}
\end{figure*}

The recovered reflectivity profiles are shown in Figure~\ref{fig:real}, along with the reflectivity profiles calculated from the sonic logs of well L-$30$ overlaid in black. Figure~\ref{fig:real} shows that inverted reflectivity for BPI and FISTA is smooth and missing details, with relatively poor amplitude recovery (following convention, red indicates positive reflectivity, and blue, negative). SBL-EM results provide a more detailed image for characterization by recovering the sparse reflectivity profiles, but the method fails to remove the noise. The {\it Nu}SPAN-$1$ and {\it Nu}SPAN-$2$ recovered reflectivity profiles show better amplitude recovery while removing the noise from the observations. Further, the {\it Nu}SPAN variants also preserve lateral continuity, especially for closely spaced interfaces, as seen around $1.1$ s on the time axis in Figure~\ref{fig:real}.

\section{Conclusions}\label{sec:conclusions}
We considered the problem of seismic reflectivity inversion based on systematically learned non-convex sparse-prior. We proposed a nonuniform sparse model based on composite regularization and solved it based on a deep-unrolled architecture for prior learning given training data. Using the proposed framework, we solved the problem of seismic reflectivity inversion, where the challenge lies in improving the resolution to characterize the subsurface. The proposed techniques outperform benchmark reconstructions given by state-of-the-art techniques on synthetic, simulated, and real datasets in terms of objective metrics.

\section*{Acknowledgments}
This work is supported by the Ministry of Earth Sciences, Government of India; Centre of Excellence in Advanced Mechanics of Materials, Indian Institute of Science (IISc), Bangalore; and Science and Engineering Research Board (SERB), India. 

\bibliography{nuspan_arxiv}
\bibliographystyle{IEEEtran}

\clearpage
\newpage

\appendices

\section*{Supplementary Material}
In this document, we provide additional experimental validation, with results on synthetic and simulated data for the proposed {\it Nu}SPAN-$1$ and {\it Nu}SPAN-$2$, in comparison with the benchmark techniques, namely, basis-pursuit inversion (BPI) \cite{chen2001atomic, zhang2011seismic}, fast iterative shrinkage-thresholding algorithm (FISTA) \cite{beck2009fast, perez2012inversion} and expectation-maximization-based sparse Bayesian learning (SBL-EM) \cite{wipf2004sparse, yuan2019seismic}.

\subsection{Experimental Results and Discussion}
In this section, we provide additional results to demonstrate the efficacy of the proposed networks, {\it Nu}SPAN-$1$ and {\it Nu}SPAN-$2$, in comparison with the benchmark techniques, BPI, FISTA, and SBL-EM, on synthetic $1$-D seismic traces and $2$-D wedge models \cite{hamlyn2014thin}, and the simulated $2$-D Marmousi$2$ model \cite{martin2006marmousi2}. We quantify the performance based on objective metrics defined in Section~\ref{metrics} of the main document, namely, Correlation Coefficient (CC), Relative Reconstruction Error (RRE), Signal-to-Reconstruction Error Ratio (SRER), and Probability of Error in Support (PES).

\subsubsection{Testing Phase -- Synthetic \texorpdfstring{$1$}{1}-D Traces}

Table~\ref{table:1d_time} shows the computational time during testing for {\it Nu}SPAN-$1$ and {\it Nu}SPAN-$2$ in comparison with the benchmark techniques, which is computed over $100$, $200$, $500$, and $1000$ test realizations of synthetic $1$-D traces. The proposed networks, namely, {\it Nu}SPAN-$1$ and {\it Nu}SPAN-$2$, require lower computational time in comparison to other techniques. FISTA, the next best to our techniques, requires two orders of magnitude more computation time than ours. Low computation times are significant in the context of reflection seismic processing, where the amount of data to be handled is large. We note that {\it Nu}SPAN-$1$ and {\it Nu}SPAN-$2$ are trained on a large number of synthetic seismic traces before testing and have longer training times, but a much shorter testing time. 

Tables~\ref{table:1d_k_0.10}, \ref{table:1d_k_0.15}, and \ref{table:1d_k_0.20} compare the performance of the proposed {\it Nu}SPAN-$1$ and {\it Nu}SPAN-$2$ with that of the benchmark techniques for sparsity factor $0.10$, $0.15$, and $0.20$. The results for {\it Nu}SPAN-$1$ and {\it Nu}SPAN-$2$ for all the tested sparsity factors are obtained by considering the model trained with sparsity factor of $0.20$, whereas the parameters of the benchmark techniques are tuned for each sparsity factor. This demonstrates the robustness of the proposed {\it Nu}SPAN to mismatch in sparsity factor, which is a significant advantage in the context of seismic reflectivity inversion, as the accurate estimation of the sparsity of the seismic reflections is challenging \cite{li2019optimal, yuan2019seismic}.

\begin{table}[t]
    \centering
    \caption{Computational time (in seconds) during testing computed over $100$, $200$ $500$, and $1000$ realizations of synthetic $1$-D traces. This comparison shows that the proposed techniques require significantly lower compute time as compared with the benchmark methods during inference. The best performance is highlighted in \textbf{boldface}. The second best performance scores are \underline{underlined}. BPI: Basis-Pursuit Inversion; FISTA: Fast Iterative Shrinkage-Thresholding Algorithm; SBL-EM: Expectation-Maximization-based Sparse Bayesian Learning; {\it Nu}SPAN-$1$ and {\it Nu}SPAN-$2$: Nonuniform Sparse Proximal Average Network -- Type $1$ and Type $2$.}
    \label{table:1d_time}
    \resizebox{0.85\columnwidth}{!}{
    \begin{tabular}{l|c|c|c|c}
        \toprule
        \multirow{2}{*}{\bfseries Method} & \multicolumn{4}{c}{\bfseries Time ($\mathrm{s}$) for \# of test realizations}\\
        \cmidrule{2-5}
         & $\mathbf{100}$ & $\mathbf{200}$ & $\mathbf{500}$ & $\mathbf{1000}$ \\
        \midrule
        BPI & ${45.0384}$ & ${93.0857}$ & ${238.8020}$ & ${503.8519}$\\
        FISTA & ${3.4676}$ & ${6.8286}$ & ${17.5604}$ & ${33.2539}$\\
        SBL-EM & ${155.0183}$ & ${313.2058}$ & ${781.2321}$ & ${1076.1833}$\\
        {\it Nu}SPAN-$1$ & $\mathbf{0.0420}$ & $\mathbf{0.0627}$ & $\mathbf{0.1505}$ & $\mathbf{0.1778}$\\
        {\it Nu}SPAN-$2$ & $\underline{0.0493}$ & $\underline{0.0707}$ & $\underline{0.1536}$ & $\underline{0.1870}$\\
        \bottomrule
    \end{tabular}}
\end{table}

\begin{table}[t]
    \centering
    \caption{Metrics averaged over $1000$ test realizations of synthetic $1$-D traces, for \textbf{sparsity factor} $\mathbf{0.10}$. {\it Nu}SPAN-$2$ shows superior amplitude recovery accuracy mentioned in terms of CC, RRE, and SRER. {\it Nu}SPAN-$1$ and {\it Nu}SPAN-$2$ are trained with sparsity factor $0.20$ and tested for sparsity factor $0.10$, whereas the benchmark techniques (BPI, FISTA, and SBL-EM) are tuned and tested for the same sparsity factor of $0.10$.}
    \label{table:1d_k_0.10}
    \resizebox{0.8\columnwidth}{!}{
    \begin{tabular}{l|c|c|c|c}
        \toprule
        \bfseries Method & \bfseries CC & \bfseries RRE & \bfseries SRER & \bfseries PES\\
        \midrule
        BPI & ${0.3764}$ & ${0.8590}$ & ${0.6749}$ & ${0.9424}$\\
        FISTA & $\underline{0.3988}$ & $\underline{0.8383}$ & $\underline{0.7770}$ & $\mathbf{0.8880}$\\
        SBL-EM & ${0.3827}$ & ${0.8520}$ & ${0.7036}$ & ${0.9424}$\\
        % {\it Nu}SPAN-$1$ & ${0.3825}$ & ${0.8820}$ & ${0.5949}$ & $\mathbf{0.7636}$\\ % k 0.05
        {\it Nu}SPAN-$1$ & ${0.3874}$ & ${0.8472}$ & ${0.7258}$ & $\underline{0.9130}$\\ % k 0.20
        {\it Nu}SPAN-$2$ & $\mathbf{0.4042}$ & $\mathbf{0.8340}$ & $\mathbf{0.7950}$ & ${0.9131}$\\
        \bottomrule
    \end{tabular}}
\end{table}

\begin{table}[t]
    \centering
    \caption{Metrics averaged over $1000$ test realizations of synthetic $1$-D traces for \textbf{sparsity factor} $\mathbf{0.15}$. The proposed {\it Nu}SPAN-$1$ and {\it Nu}SPAN-$2$ show superior amplitude and support recovery, although they are trained with sparsity factor $0.20$ but tested on a different sparsity factor. The best performance is highlighted in \textbf{boldface}. The second best performance scores are \underline{underlined}.}
    \label{table:1d_k_0.15}
    \resizebox{0.8\columnwidth}{!}{
    \begin{tabular}{l|c|c|c|c}
        \toprule
        \bfseries Method & \bfseries CC & \bfseries RRE & \bfseries SRER & \bfseries PES\\
        \midrule
        BPI & ${0.3602}$ & ${0.8693}$ & ${0.6126}$ & ${0.9155}$\\
        FISTA & ${0.3727}$ & ${0.8585}$ & ${0.6679}$ & ${0.8759}$\\
        SBL-EM & ${0.3646}$ & ${0.8658}$ & ${0.6299}$ & ${0.9155}$\\
        % {\it Nu}SPAN-$1$ & ${0.3047}$ & ${0.9832}$ & ${0.1093}$ & $\mathbf{0.7724}$\\ % k 0.05
        {\it Nu}SPAN-$1$ & $\underline{0.3804}$ & $\underline{0.8528}$ & $\underline{0.6965}$ & $\mathbf{0.8725}$\\ % k 0.20
        {\it Nu}SPAN-$2$ & $\mathbf{0.3904}$ & $\mathbf{0.8448}$ & $\mathbf{0.7377}$ & $\underline{0.8726}$\\
        \bottomrule
    \end{tabular}}
\end{table}

\begin{table}[t]
    \centering
    \caption{Metrics averaged over $1000$ test realizations of synthetic $1$-D traces for \textbf{sparsity factor} $\mathbf{0.20}$. {\it Nu}SPAN-$1$ and {\it Nu}SPAN-$2$ outperform the benchmark techniques in both amplitude and support recovery. Here, all the methods are tuned/trained with the same sparsity factor as they are tested for.}
    \label{table:1d_k_0.20}
    \resizebox{0.8\columnwidth}{!}{
    \begin{tabular}{l|c|c|c|c}
        \toprule
        \bfseries Method & \bfseries CC & \bfseries RRE & \bfseries SRER & \bfseries PES\\
        \midrule
        BPI & ${0.3550}$ & ${0.8735}$ & ${0.5910}$ & ${0.8895}$\\
        FISTA & ${0.3619}$ & ${0.8687}$ & ${0.6144}$ & ${0.8781}$\\
        SBL-EM & ${0.3240}$ & ${0.8999}$ & ${0.4602}$ & ${0.8899}$\\
        % {\it Nu}SPAN-$1$ & ${0.2709}$ & ${1.0698}$ & ${-0.1984}$ & $\mathbf{0.7592}$\\ % k 0.05
        {\it Nu}SPAN-$1$ & $\underline{0.3750}$ & $\underline{0.8574}$ & $\underline{0.6724}$ & $\mathbf{0.8339}$\\ % k 0.20
        {\it Nu}SPAN-$2$ & $\mathbf{0.3812}$ & $\mathbf{0.8523}$ & $\mathbf{0.6984}$ & $\underline{0.8340}$\\
        \bottomrule
    \end{tabular}}
\end{table}

\subsubsection{Testing Phase -- Synthetic \texorpdfstring{$2$}{2}-D Wedge Models}

\begin{figure*}[t]
    \centering
    \resizebox{0.6\linewidth}{!}{
    \includegraphics{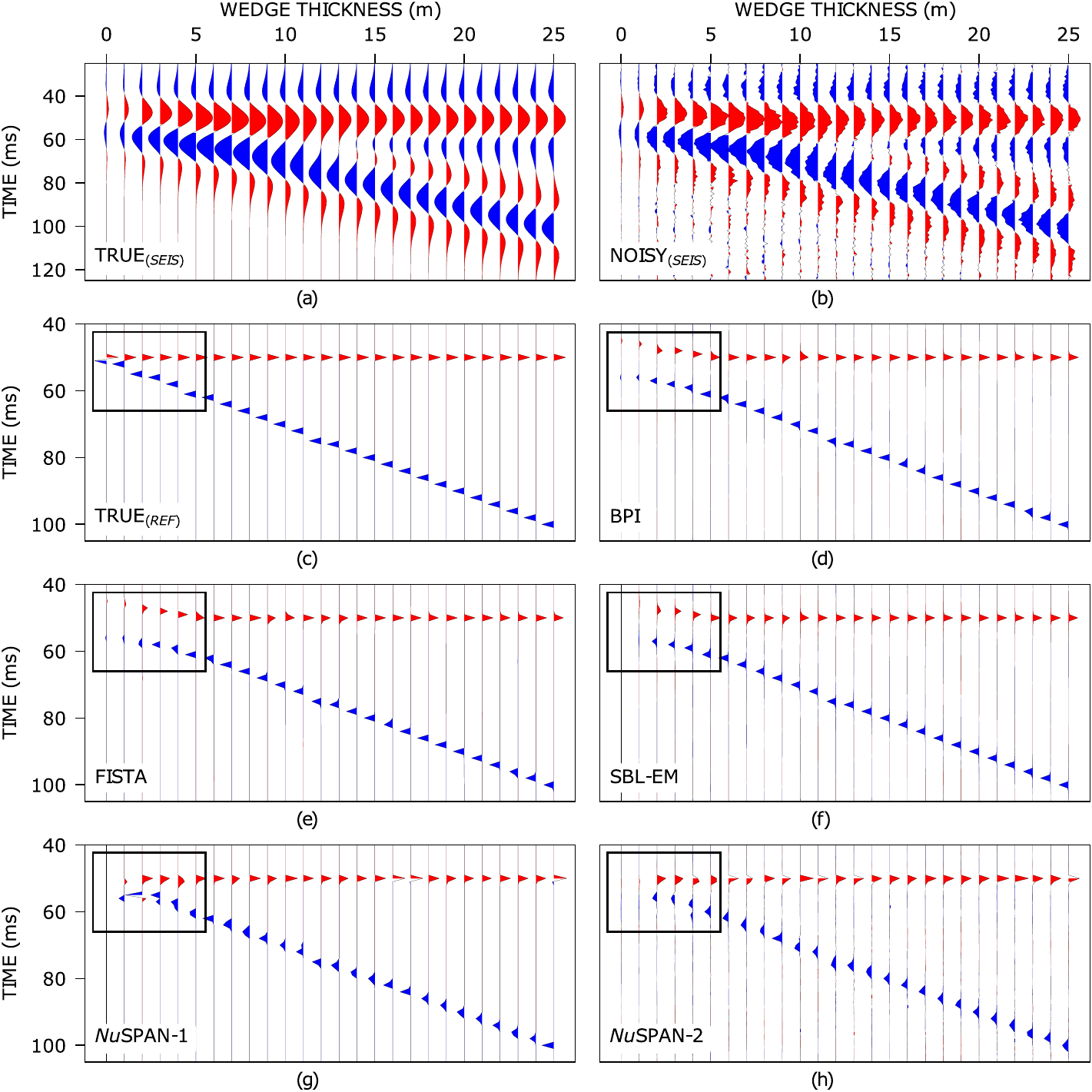}}
    \caption{\protect\rgbsymbol~Results for a synthetic $2$-D odd (PN) wedge model. True (a) seismic traces and (c) reflectivity; (b) Noisy seismic traces with input SNR $10$ dB. (d)-(f) Recovered reflectivity signatures for BPI, FISTA, and SBL-EM show that these methods fail to resolve reflectors with wedge thickness $<~5$ m, evident from the diverging interfaces highlighted by the rectangle in black. (g)-(h) The recovered reflectivity profiles for {\it Nu}SPAN-$1$ and {\it Nu}SPAN-$2$ show better resolution of reflector locations at wedge thickness $<~5$ m.}
    \label{fig:wedge_pn}
\end{figure*}

\begin{figure*}[t]
    \centering
    \resizebox{0.7\linewidth}{!}{
    \includegraphics{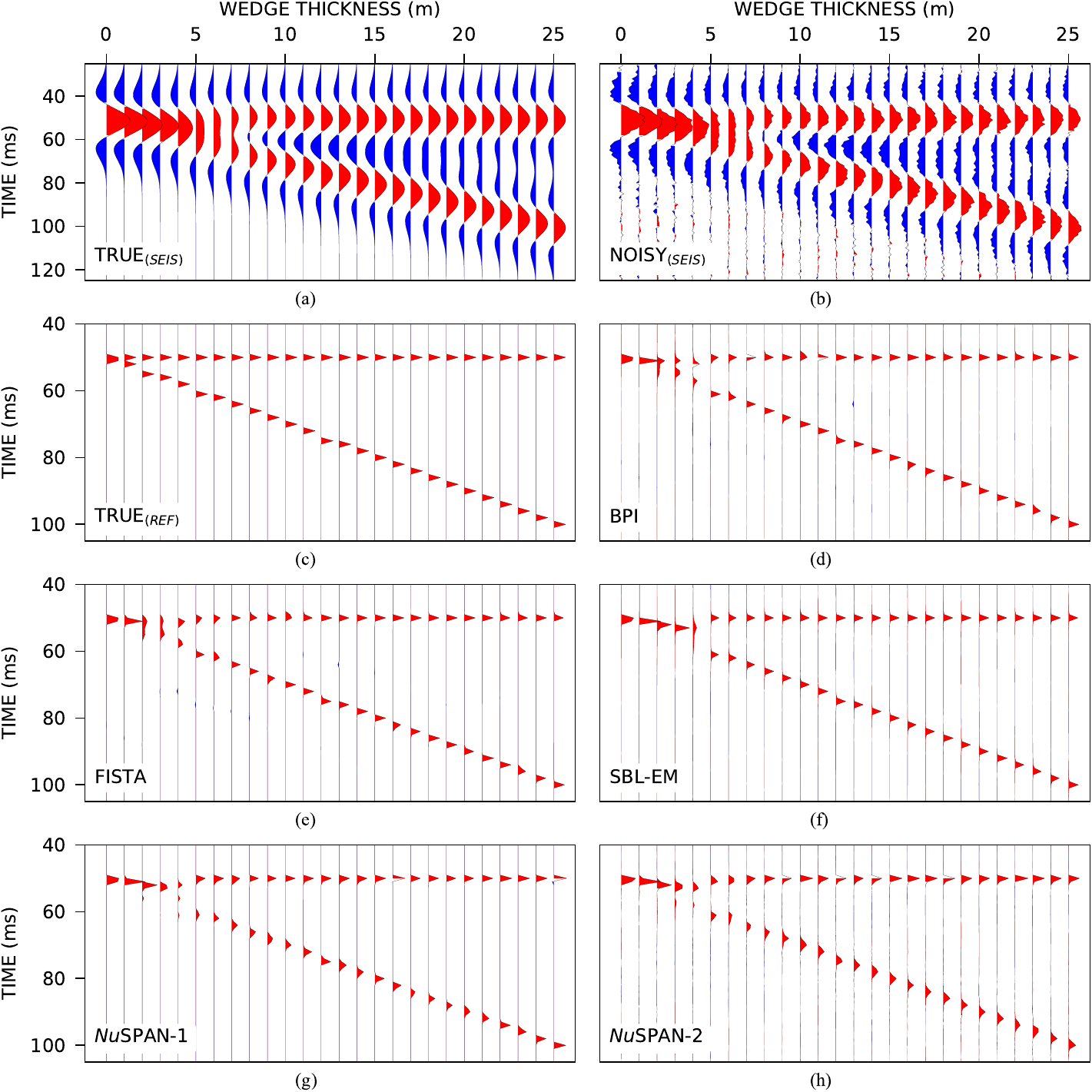}}
    \caption{\protect\rgbsymbol~Results for a synthetic $2$-D even (PP) wedge model. True (a) seismic traces and (c) reflectivity; (b) Noisy seismic traces with input SNR $10$ dB; (d)-(h) Recovered reflectivity. All methods resolve the even wedge model well. {\it Nu}SPAN-$1$ does not introduce spurious supports as the other methods, seen as low-amplitude spikes in regions away from the two interfaces in the results of the other methods.}
    \label{fig:wedge_pp}
\end{figure*}

In Section~\ref{subsubsec:results_wedge} of the main document, we have shown the results for one odd (NP) and one even (NN) wedge model, where N and P denote the polarity, negative and positive, respectively, of the two interfaces of the wedge models. Here, we present results for two wedge model variants, one odd (PN -- Table~\ref{table:pn} and Figure~\ref{fig:wedge_pn}) and one even (PP -- Table~\ref{table:pp} and Figure~\ref{fig:wedge_pp}).

\begin{table}[t]
    \centering
    \caption{Results for a synthetic $2$-D odd (PN) wedge model. {\it Nu}SPAN-$1$ shows superior amplitude recovery in terms of CC, and outperforms the benchmark techniques in support recovery quantified in terms of PES.}
    \label{table:pn}
    \resizebox{0.8\columnwidth}{!}{
    \begin{tabular}{l|c|c|c|c}
        \toprule
        \bfseries Method & \bfseries CC & \bfseries RRE & \bfseries SRER & \bfseries PES\\
        \midrule
        BPI & ${0.8002}$ & ${0.2711}$ & $\mathbf{13.0565}$ & ${0.9933}$\\
        FISTA & ${0.7984}$ & $\underline{0.2684}$ & $\underline{12.8473}$ & $\underline{0.7058}$\\
        SBL-EM & $\underline{0.8096}$ & $\mathbf{0.2622}$ & ${10.3299}$ & ${0.9933}$\\
        {\it Nu}SPAN-$1$ & $\mathbf{0.8650}$ & ${0.2695}$ & ${8.1181}$ & $\mathbf{0.6019}$\\
        {\it Nu}SPAN-$2$ & ${0.7877}$ & ${0.3440}$ & ${5.3725}$ & ${0.9896}$\\
        \bottomrule
    \end{tabular}}
\end{table}

\begin{table}[t]
    \centering
    \caption{Results for a synthetic $2$-D even (PP) wedge model. {\it Nu}SPAN-$1$ outperforms the benchmark techniques in support recovery in terms of PES, whereas BPI, FISTA, and SBL-EM show better amplitude recovery accuracy mentioned in terms of CC, RRE, and SRER.}
    \label{table:pp}
    \resizebox{0.8\columnwidth}{!}{
    \begin{tabular}{l|c|c|c|c}
        \toprule
        \bfseries Method & \bfseries CC & \bfseries RRE & \bfseries SRER & \bfseries PES\\
        \midrule
        BPI & $\underline{0.8586}$ & $\underline{0.2608}$ & $\mathbf{12.9680}$ & ${0.9933}$\\
        FISTA & $\mathbf{0.8791}$ & $\mathbf{0.2274}$ & $\underline{12.3515}$ & $\underline{0.6673}$\\
        SBL-EM & ${0.8491}$ & ${0.3205}$ & ${10.559}$ & ${0.9933}$\\
        {\it Nu}SPAN-$1$ & ${0.8037}$ & ${0.3689}$ & ${8.6339}$ & $\mathbf{0.5680}$\\
        {\it Nu}SPAN-$2$ & ${0.7516}$ & ${0.4540}$ & ${5.45024}$ & ${0.9896}$\\
        \bottomrule
    \end{tabular}}
\end{table}

\subsubsection{Testing Phase -- Simulated Marmousi\texorpdfstring{$2$}{2} model}

Initial evaluations on the Marmousi$2$ model \cite{martin2006marmousi2} showed very low PES for BPI and SBL-EM (Table~\ref{table:mm_unmute}), possibly due to the spurious support estimates introduced by these methods complementing the low-amplitude spikes ($<~1\%$ of the absolute of the maximum amplitude) in the Marmousi$2$ model. When these spikes were muted, the CC, RRE, and SRER for all the methods improved, but the PES, now evaluated only on significant interfaces, was much higher for BPI and SBL-EM, and lower for {\it Nu}SPAN-$1$ (Table~\ref{table:mm_mute} in the main document).

In Table~\ref{table:mm_unmute}, we also provide the training and testing times of the benchmark methods, namely, BPI, FISTA and SBL-EM, and the proposed {\it Nu}SPAN-$1$ and {\it Nu}SPAN-$2$, on the complete Marmousi$2$ model. Comparing the three learning-based methods, i.e., SBL-EM, {\it Nu}SPAN-$1$, and {\it Nu}SPAN-$2$, shows that the proposed approaches have lower combined training and testing time than SBL-EM. Although the combined time for FISTA is low, {\it Nu}SPAN-$2$ achieves higher accuracy in terms of both amplitude and support recovery after training on a larger synthetic dataset.

\begin{table*}[t]
    \centering
    \caption{Results for a portion of the Marmousi$2$ model corresponding to Table~\ref{table:mm_mute} and Figure~\ref{fig:mm} in Section~\ref{results:marmousi2} of the main document, without muting low-amplitude spikes. {\it Nu}SPAN-$2$ provides the best amplitude recovery. The spurious supports introduced by BPI and SBL-EM coincide with some of the low-amplitude spikes, leading to very low PES values. The reported times are for the complete model (training times on one RTX 2080 Ti GPU). Training times are applicable only to {\it Nu}SPAN-$1$ and {\it Nu}SPAN-$2$. The best performance is highlighted in \textbf{boldface}. The second best performance scores are \underline{underlined}.}
    \label{table:mm_unmute}
    \resizebox{.8\linewidth}{!}{
    \begin{tabular}{l|c|c|c|c||c|c|c}
        \toprule
        \multirow{2}{*}{\bfseries Method} & \multirow{2}{*}{\bfseries CC} & \multirow{2}{*}{\bfseries RRE} & \multirow{2}{*}{\bfseries SRER} & \multirow{2}{*}{\bfseries PES} & \multicolumn{3}{c}{\bfseries Time ($\mathrm{h}$)}\\
        \cmidrule{6-8}
         &  &  &  &  & \bfseries Training & \bfseries Testing & \bfseries Training + Testing\\
        \midrule
        BPI & ${0.9472}$ & ${0.0878}$ & ${14.7626}$ & $\mathbf{0.0181}$ & - & $16.2671$ & ${16.2671}$ \\% 16.2671 hr (58561.5337 s)
        FISTA & ${0.9406}$ & ${0.1021}$ & ${13.8227}$ & ${0.912}$ & - & $5.5636$ & $\mathbf{5.5636}$ \\% 5.5636 hr (20028.9425 s)
        % FISTA & ${0.9397}$ & ${0.1033}$ & ${13.8201}$ & ${0.9119}$ & - & $5.5636$ & $\mathbf{5.5636}$ \\% 5.5636 hr (20028.9425 s)
        SBL-EM & $\underline{0.9548}$ & $\mathbf{0.0686}$ & $\mathbf{18.1117}$ & $\mathbf{0.0181}$ & - & $101.2659$ & $101.2659$ \\% 101.2659 Hrs (364557.194 s)
        {\it Nu}SPAN-$1$ & ${0.9375}$ & ${0.1035}$ & ${14.2261}$ & ${0.9650}$ & $\underline{9.8069}$ & $\mathbf{0.0737}$ & $9.8806$ \\% 0.0737 hr (265.3918 s)
        {\it Nu}SPAN-$2$ & $\mathbf{0.9590}$ & $\underline{0.0718}$ & $\underline{15.0579}$ & $\underline{0.2021}$ & $\mathbf{8.6007}
        $ & $\underline{0.1024}$ & $\underline{8.7031}$ \\% 0.8206 hr (2954.0337 s)
        \bottomrule
    \end{tabular}}
\end{table*}

\end{document}